\documentclass[onecolumn,sort&compress,numbers]{els-mrw} 

\usepackage{amsmath,amssymb,amsfonts,amsthm,makeidx,graphicx}
\usepackage{txfonts}
\usepackage{helvet}
\usepackage[hyperindex,breaklinks]{hyperref}
\usepackage{epstopdf}
\usepackage{mathtools}
\usepackage{enumitem}

\begin{document}

\newcommand{\nc}{\newcommand}
\nc{\be}{\begin{equation}}
\nc{\ee}{\end{equation}}
\nc{\ba}{\begin{eqnarray}}
\nc{\ea}{\end{eqnarray}}
\nc{\bi}[1]{\bibitem{#1}}
\nc{\al}{\alpha}
\nc{\de}{\delta}
\nc{\ep}{\epsilon}
\nc{\ze}{\zeta}
\nc{\et}{\eta}
\renewcommand{\th}{\theta}
\nc{\Th}{\Theta}
\nc{\ka}{\kappa}
\nc{\rh}{\rho}
\nc{\si}{\sigma}
\nc{\ta}{\tau}
\nc{\up}{\upsilon}
\nc{\ph}{\phi}
\nc{\ch}{\chi}
\nc{\ps}{\psi}
\nc{\om}{\omega}
\nc{\Ga}{\Gamma}
\nc{\De}{\Delta}
\nc{\La}{\Lambda}
\nc{\Si}{\Sigma}
\nc{\Up}{\Upsilon}
\nc{\Ph}{\Phi}
\nc{\Ps}{\Psi}
\nc{\Om}{\Omega}
\nc{\ptl}{\partial}
\nc{\del}{\nabla}
\nc{\ov}{\overline}
\nc{\fr}{\frac}
\nc{\ga}{\mathrel{\raise.3ex\hbox{$>$\kern-.75em\lower1ex\hbox{$\sim$}}}}
\nc{\la}{\mathrel{\raise.3ex\hbox{$<$\kern-.75em\lower1ex\hbox{$\sim$}}}}

\def\lsim{\mathrel{\rlap{\lower4pt\hbox{\hskip1pt$\sim$}}
    \raise1pt\hbox{$<$}}}                
\def\gsim{\mathrel{\rlap{\lower4pt\hbox{\hskip1pt$\sim$}}
    \raise1pt\hbox{$>$}}}                

\newcommand{\AR}[1]{\color{blue} [AR:#1] \color{black}}


\chapter{Electric Dipole Moments and New Physics}\label{chap1}

\author[1]{Maxim Pospelov}%
\author[2]{Adam Ritz}%

\address[1]{\orgname{University of Minnesota}, \orgdiv{William I. Fine Theoretical Physics Institute, School of Physics and Astronomy}, \orgaddress{Minneapolis, MN 55455, USA}}
\address[2]{\orgname{University of Victoria}, \orgdiv{Department of Physics and Astronomy}, \orgaddress{Victoria, BC V8P 5C2, Canada}}


\maketitle

\begin{abstract}[Abstract]
	Searches for intrinsic electric dipole moments (EDMs) of nucleons, atoms and molecules are precision flavor-diagonal 
probes of new $CP$-odd physics, as motivated by the need to explain the matter-antimatter asymmetry in the universe. We review and summarise the effective field theory analysis of the observable
EDMs in terms of a general set of $CP$-odd operators at 1~GeV, and the ensuing model-independent constraints on new physics. We also review and discuss the EDMs induced by $CP$-violation in the Standard Model, and the implications of EDM limits for various models of physics beyond the Standard Model.
\end{abstract}

\begin{keywords}
 	electric dipole moments\sep $CP$ violation\sep baryon asymmetry\sep new physics
\end{keywords}


\begin{glossary}[Nomenclature]
	\begin{tabular}{@{}lp{34pc}@{}}
		EDM & Electric Dipole Moment\\
        $CP$ & Charge-Parity\\
        BSM & Beyond the Standard Model\\
	\end{tabular}
\end{glossary}

\section*{Objectives}
\begin{itemize}
	\item To understand the origin of fundamental Dirac fermion EDMs as $T$ and $P$-odd (and thus $CP$-odd) quantities. 
    \item To describe the mechanisms that generate EDM observables in nucleons, nuclei, atoms and molecules from underlying $CP$-odd sources, and the physics responsible for various enhanements.
    \item To summarize the contributions of $CP$-violation in the Standard Model from the CKM phase to observable EDMs in various categories of observational interest.
	\item To highlight the use of EDM measurements to probe new fundamental sources of $CP$-violation in physics beyond the Standard Model.
\end{itemize}


\section{Introduction}

The Standard Model (SM) of particles and fields is extremely successful in describing the majority of  phenomena in fundamental subatomic physics. With the discovery of the Higgs boson in 2012 at the Large Hadron Collider (LHC), and with its properties so far consistent with the most minimal version of the SM \cite{ATLAS:2022vkf,CMS:2022dwd}, there is now evidence for all the essential fields and their most important interactions in the SM Lagrangian.\footnote{All but one, the $\theta$-term of QCD.} Nonetheless, beyond more fundamental questions about the nature of short distance physics, there are still compelling empirical problems that the SM is seemingly unable to explain, such as the nature of dark matter, the origin of the matter-antimatter asymmetry, and indeed the nature of neutrino mass. This provides ample motivation to search for physics beyond the Standard Model (BSM). Moreover, with barriers to substantially raising the energy scale of new colliders due to cost, time and technology, indirect searches for BSM physics continue to acquire a more prominent role. In this chapter, we review an important class of observables in indirect searches for BSM physics, namely electric dipole moments (EDMs).\footnote{In addition to this compact exposition of EDM-related phenomena, more extensive reference material can be found in several earlier reviews \cite{Khriplovich:1997ga,Ginges:2003qt,Pospelov:2005pr,Engel:2013lsa}.}

An important theme among the strong empirical hints for new physics noted above is that they do not point directly to a new characteristic BSM energy scale. This scale could reside at or above LHC energies, or indeed be very light if the new physics is sufficiently weakly coupled to the SM. In particular, the phenomenon of neutrino oscillations establishes the need for SM extensions that can accommodate non-zero neutrino masses and mixing angles. This new physics could be very light (Dirac neutrino masses) or very heavy in the minimal seesaw model for Majorana neutrino mass. The presence of dark matter (DM) in the Universe, seen through its gravitational interactions at distance scales of kiloparsecs and larger, points to the existence of additional sources of matter that can cluster and be non-relativistic prior to the epoch of matter-radiation equality. However, the universality of gravitational interactions makes it difficult to pinpoint the mass scale associated with dark matter particles. New physics is also hinted at by the lack of evidence for the $\theta$-term in the QCD, and thus the fine-tuned value of $\theta_{\rm QCD}$ (dubbed the strong $CP$ problem), possibly implying a dynamical mechanism for its relaxation with an associated light pseudo-Nambu-Goldstone boson, {\em i.e.} the QCD axion. 

The landscape of physically realizable BSM scenarios can be conveniently represented using an effective field theory (EFT) Lagrangian normalized at the electroweak scale. In addition to the SM spinor, gauge and Higgs/scalar fields, denoted schematically as $(\psi, A, H)$, this EFT can include the effects of heavy BSM physics which is integrated out and encapsulated by effective SM operators of increasing dimension, and by the existence of interactions with light BSM fields that on account of their weak coupling to the SM comprise a ``dark sector" (DS):
\begin{align}
    {\cal L}_{\rm EFT} =& m_H^2H^\dagger H + {\cal L}_{\rm SM,\, dim\, 4}(A,\psi,H) +\sum_{n\geq 5} 
\frac{1}{\Lambda^{n-4}}{\cal O}^{(n)}_{{\rm SM,\,dim\,}n}(A,\psi,H)
\nonumber\\
&+ {\cal L}_{\rm portals}(\psi, A, H; \varphi_{\rm DS})+ {\cal L}_{DS}(\varphi_{\rm DS}).
\label{eq1}
\end{align}
The first line of this EFT Lagrangian includes the unique $dim=2$ operator, the Higgs mass, and all possible $dim=4$ operators of the SM respecting the $SU(2)\times U(1)\times SU(3)$ gauge symmetry for the three generations of quarks and leptons. The final term on the first line includes possible extensions by higher-dimensional operators generated by BSM at a new scale $\Lambda$ that is taken to be large compared to the SM electroweak scale. The existence of these operators ${\cal O}^{(n)}_{{\rm SM,\,dim\,}n}$ can be probed in a variety of ways. Famously, the only dimension five operators are the Weinberg operators generating a Majorana mass for the SM neutrinos \cite{Weinberg:1979sa}. The multitude of dimension six and higher operators can be explored via high-energy processes, observing the change in the predicted cross sections, or via precision measurements at low energies provided such changes can be distinguished from SM effects. The second line of Eq.~{\ref{eq1}) allows for new DS fields (denoted schematically as $\varphi_{\rm DS})$ that are light relative to the electroweak scale, and the so-called portal couplings, contained in ${\cal L}_{\rm portals}$, that connect the SM with the DS while respecting all relevant SM symmetries. Since these new degrees of freedom are by definition light relative to the weak scale, they can be probed both on-shell at colliders and off-shell using low-energy indirect strategies. Motivated in part by the need to explain dark matter, there has been extensive investigation of dark sectors and portal couplings over the past decade.

Historically, tests of fundamental symmetries have provided some of our most powerful probes of physics beyond the Standard Model (SM). For example, the (perturbatively) conserved baryon and lepton numbers are a consequence of {\em accidental} symmetries, and can be violated at the $dim=6$ level by a four-fermion operator involving three quark fields and one lepton field. The corresponding scale probed by experimental searches for proton decay reaches $\Lambda \propto 10^{16}$\,GeV, a scale that is well beyond the foreseeable reach of accelerators. In this article we will discuss the consequences of testing time reversal symmetry $T$ in flavor-conserving processes using electric dipole moments (EDMs) via precision measurements involving neutrons, heavy atoms and molecules. The focus on $T$ symmetry (and thus $CP$ symmetry according to the $CPT$ theorem) is not accidental. The SM incorporates the violation of $C$ (charge conjugation) and  $P$ (parity) in a maximal way at the electroweak scale through the $V-A$ structure of the weak couplings. The combined $CP$ symmetry is also violated in the SM, but at a much more subtle level. The only measured source of $CP$-violation is in the quark sector of the electroweak theory and relies on the opportunity for a $CP$-odd phase to enter nontrivial quark flavour mixing \cite{Kobayashi:1973fv}. In particular, all modes of $CP$ violation observed
thus far can be consistently described by the single (physical) phase in the unitary three-generation Cabibbo-Kobayashi-Maskawa (CKM) quark mixing matrix $V$. Its strength is
characterized by the so-called Jarlskog invariant, ${\cal J}={\rm Im}[V_{us} V_{cd} V_{cs}^* V_{ub}^*] \sim 3 \times 10^{-5}$ \cite{Jarlskog:1985ht}, and in practice is often suppressed further by the Yukawa structure of the SM. Under very general assumptions of Lorentz invariance, locality and spin-statistics, $CPT$ is identically conserved, and thus any violation of $CP$ implies violation of time-reversal symmetry $T$. The CKM phase therefore generates irreducible EDM contributions, but in practice their predicted values are exceedingly small, suppressed not only by $J$, but also by a multitude of loop factors and by at least two powers of the Fermi constant $G_F$. This suppression of the SM contribution presents an opportunity to use EDM observables as highly sensitive probes of BSM $CP$-violation in current and future generation of EDM experiments.  

To define the concept of a fundamental electric dipole moment, consider a particle at rest and follow its spin degrees of freedom as they interact with background electric and magnetic fields:
\be
 H = - \mu {\bf B} \cdot \frac{\bf S}{S} - d {\bf E}\cdot  \frac{\bf S}{S} ,
\label{deff}
\ee
where $H$ is the interaction Hamiltonian. The magnetic moment $\mu$ conserves all discrete symmetries because the spin operator and magnetic field have exactly the same transformation properties under $P$ and $T$. The quantity $d$, the EDM, breaks both $P$ and $T$, and is therefore expected to be small. The presence of $d$ will induce an analogue of linear Zeeman-type energy level splitting, induced by an external electric field, as was first emphasized by Purcell and Ramsey \cite{PhysRev.78.807}. 

It is straightforward to discern how an EDM $d$ can descend from the electroweak-scale effective Lagrangian (\ref{eq1}). Consider, for example, the EDM of an electron. The simple single-particle Hamiltonian (\ref{deff}) can be thought of as originating from an effective $CP$-odd dimension five operator in the QED Lagrangian: $(-id/2)\times  \bar\psi \sigma_{\mu\nu}\gamma_5 F_{\mu\nu}\psi$, where $\psi$ is the Dirac spinor field associatd with the electron and $F_{\mu\nu}$ is the electromagentic field strength. Recalling, however, that the left- and right-handed spinor fields of SM leptons belong to different representations and that their coupling necessitates a Higgs field insertion \cite{Weinberg:1967tq}, this operator can br understood as descending from $dim=6$ operators of the following form: 
\begin{align}
   {\cal L}_{\rm EFT} &\sim \frac{1}{\Lambda_1^2} e^{i\phi_1} H\bar E  \sigma_{\mu\nu} B_{\mu\nu} L + \frac{1 }{\Lambda_2^2} e^{i\phi_2}H\bar E  \sigma_{\mu\nu} \tau^a W^a_{\mu\nu} L +(h.c.),
\end{align}
where $L,H$ refer to the electroweak lepton and Higgs field doublets, $E$ is the (right-handed) electroweak lepton singlet, and $B$ and $W^a$ are the $U(1)$ and $SU(2)$ gauge fields. The important scaling that emerges from this simple analysis is that the EDM will be inversely proportional to the {\em square} of the new physics scale, 
\begin{eqnarray}
\label{naive}
    d\propto \sin\ph \frac{\langle H\rangle}{\Lambda_{\rm BSM}^2}.
\end{eqnarray}
Comparing this scaling directly with the existing experimental sensitivity (exhibited for several characteristic examples in Table.~\ref{explimit}), one may conclude that if the $CP$-violating phase $\phi\sim O(1)$, the {\em maximum} energy scales probed by EDMs can be as high as $\Lambda_{\rm BSM} \sim 10^9$\,GeV. However, in the majority of BSM examples the estimate (\ref{naive}) must further account for the facts that $CP$-violating phases can be small, and that coefficients are typically suppressed by loop factors and the small Yukawa couplings of light fermions. Even with these caveats, the scale of new physics probed by EDMs can be much larger than the reach of current colliders. 

This line of argument, connecting EDMs to high-scale new physics is of course not a universal one. BSM physics in a dark sector could be relatively light and still induce EDMs. Moreover, the interpretation of EDMs in the hadornic sector is complicated by the theoretical realization that the non-perturbative QCD solution to the so-called $U(1)_A$-problem renders the QCD $\theta$ angle an observable quantity \cite{tHooft:1976rip}. EDMs may therefore contain contributions from QCD scales, mediated by $\theta_{\rm QCD}$. At the same time, since the $\theta$-term appears in the dimension four Standard Model Langrangian, finite corrections to $\theta_{\rm QCD}$ may in turn arise from arbitrarily high energy scales. 

In addition to probing generic BSM extensions of the SM, there is also a strong empirical motivation to search for new sources of $CP$-violation which could provide a dynamical explanation for
the baryon  asymmetry in the Universe. The Sakharov criteria \cite{Sakharov:1967dj} require $C$ and $CP$ violation for successful baryogenesis and, while the 
SM itself does violate these symmetries, it apparently fails by many orders of magnitude in explaining the magnitude of the observed baryon to photon ratio $\eta_b/s \sim 10^{-10}$. While there are several promising scenarios for how an appreciable baryon asymmetry can be generated dynamically, experimental evidence is lacking. As EDMs are significantly induced in some scenarios ({\em e.g.} electroweak baryogenesis \cite{Cohen:1993nk}), while remaining small in others ({\em e.g.} baryogenesis via leptogenesis \cite{Fukugita:1986hr}), they can provide a discriminator between different mechanisms, and therefore are important for our understanding of cosmological history.

Searches for intrinsic EDMs have a long history, stretching back to the pioneering work of Purcell and Ramsey \cite{PhysRev.78.807} who
first used the neutron EDM to probe the conservation of parity in strong interactions.
After many decades of progress and innovation,  several classes of EDM-related observables have been identified as the most competitive:
\begin{enumerate}[leftmargin=.5in]
    \item Neutron EDM
    \item EDMs in atoms and molecules with closed electron shells test the coupling of the electric field to the nuclear spin, and We will refer to this class as diamagnetic EDMs. 
    \item EDMs in atoms and molecules with uncompensated electron spins are uniquely sensitive to electron EDMs, and we will refer to them as paramagnetic EDMs. 
\end{enumerate}
No fundamental EDM has yet been observed and the most sensitive EDM limits, expressed in terms of the most convenient $CP$-odd quantities to be discussed further in the next section, are shown in Table 1. In addition, there are a variety of new proposals and experimental efforts underway that promise to elevate sensitivity to new levels in the near future. Indeed, as experimental observables, EDMs have a number of special features which can enhance their value as indirect probes of new physics: 
\begin{itemize}
    \item As amplitude-level observables, their dependence on new physics scales is enhanced relative to cross sections. This feature is shared, for example, with anomalous magnetic moments.
    \item The absence of a significant SM CKM background for EDM searches (leaving aside the question of why the QCD $\theta$-angle is small) places EDM observables in the same category as other nearly background-free measurements. These include nucleon stability tests ({\em e.g.} proton decay, di-nucleon annihilation etc), and charged lepton flavor conservation ($\mu\to e\gamma$, $\mu-e$ conversion, etc).
    \item The utility of atomic and molecular systems for EDM measurements presents novel opportunities to enhance sensitivity, e.g. through macroscopic effects such as polarization that magnify the internal electric field, and the possibility of an extended coherence time. Schematically, we can write
    \begin{equation}
     {\rm EDM\, observable}\,\propto {\cal P}_{\rm system}(E,\ldots)\times \sin(\ph)
    \end{equation}
    where $\sin(\phi)$ is the $CP$-odd source, and ${\cal P}_{\rm system}$ reflects the nonlinear response of the system to external parameters such as the electric field, which can be engineered to be very large, e.g. in molecular states. This provides a cost-effective means of dramatically improving sensitivity, one that is not available for other probes where improvements depend more directly on system size (e.g. proton decay), or source intensity (e.g. lepton flavour violation).
\end{itemize}

The rest of the chapter is organized as follows. In Section~\ref{Sec2}, we summarize how the EDM constraints can be translated to a set of induced bounds on the most generic class of $CP$-odd operators normalized at 1 GeV \cite{Pospelov:2005pr}. This procedure clarifies how various enhancement and suppression factors in atomic and molecular systems impact the ultimate sensitivity of those measurements to underlying sources of $CP$-violation when applied to constrain models of new physics. In Section~\ref{Sec3}, we review the contributions to these EDMs from the SM sources of $CP$-violation. In Section~\ref{Sec4}, we turn to constraints on models of new physics, discussing some applications to physics at the TeV scale and above. We also discuss the implications of the strong $CP$ problem and study the connection between EDMs and dark sectors. We finish with some concluding remarks in Section~\ref{Sec5}.

Throughout this review, we consistently use natural units: $\hbar=c=1$, so that all mass (or inverse length) scales can be expressed in terms of electron volts (eV).

\begin{table}
\begin{center}
\begin{tabular}{|c|c|c|}
 \hline
  Class & EDM & Current Bound  \\
  \hline
  Paramagnetic & HfF$^+$ & $|d_e^{\rm equiv}| < 4.1 \times 10^{-30} e\, {\rm cm}$ \cite{Roussy:2022cmp}  \\
 & ThO & $|d_e^{\rm equiv}| < 1.1 \times 10^{-29} e\, {\rm cm}$ \cite{ACME:2018yjb}
  \\
  Diamagnetic & $^{199}$Hg & $|d_{\rm Hg}| < 7.4   \times 10^{-30}  e\, {\rm cm}$ \cite{Graner:2016ses}  \\
  Nucleon & $n$ & $|d_n|  <  1.8\times 10^{-26} e\, {\rm cm}$ \cite{Abel:2020pzs} \\
  \hline
\end{tabular}
\caption{Leading direct constraints within three representative classes of EDM observables. See the discussion in Section 2 for further details of the paramagnetic and diamagnetic categories and the notation.}
\end{center}
\label{explimit}
\end{table}

\section{$CP$-odd operators and electric dipole moments}
\label{Sec2}

In this section we will briefly review the relevant scales in the effective field theory hierarchy, relating the observable EDMs to
$CP$-odd operators at various convenient physical scales. Our starting point will be an effective Lagrangian normalized at just above the hadronic scale of 1 GeV. The $CP$-odd operators in (\ref{eq1}) formulated just above the electroweak (EW) scale can be evolved down to this benchmark scale, using perturbative renormalization group evolution. In the process, some explicit features of $SU(2)\times U(1)$ invariance are obscured, and $\langle H\rangle $ is subsumed into redefined Wilson coefficients.  Including the most significant flavor-diagonal $CP$-odd operators
(see e.g. \cite{Pospelov:2005pr}) up to dimension six, the corresponding $CP$-odd effective Lagrangian takes the form,
\begin{eqnarray}
{\cal{L}}^{\rm CP}_{\rm eff,1\,GeV} &=& \frac{\bar{\theta}g_s^2}{32\pi^{2}} G^{a}_{\mu\nu} \widetilde{G}^{\mu\nu , a} -
\frac{i}{2} \sum_{i=e,\mu,u,d,s} d_i\ \overline{\psi}_i (F\sigma)\gamma_5 \psi_i  -
\frac{i}{2} \sum_{i=u,d,s} \widetilde{d}_i\ \overline{\psi}_i g_s (G\sigma)\gamma_5\psi_i \nonumber\\
  && + \frac{1}{3} w\  f^{a b c} G^{a}_{\mu\nu} \widetilde{G}^{\nu \beta , b}
G^{~~ \mu , c}_{\beta} + \sum_{i,j=e,\mu,q} C_{ij} (\bar{\ps}_i \ps_i) (\bar \ps_j i\gamma_5 \ps_j)+ \cdots \label{leff}
\end{eqnarray}
In this expression, $F$ and $G$ are the electromagnetic and gluon field strength tensors respectively, while $\ps_i$ denotes the
Dirac fields of the various quarks and leptons with masses below the QCD scale. The quark content therefore includes at most up, down and strange flavours, $q=u,d,s$, with heavier quarks having been integrated out. 

\begin{figure}[t]
	\centering
	\includegraphics[angle=270,width=8cm]{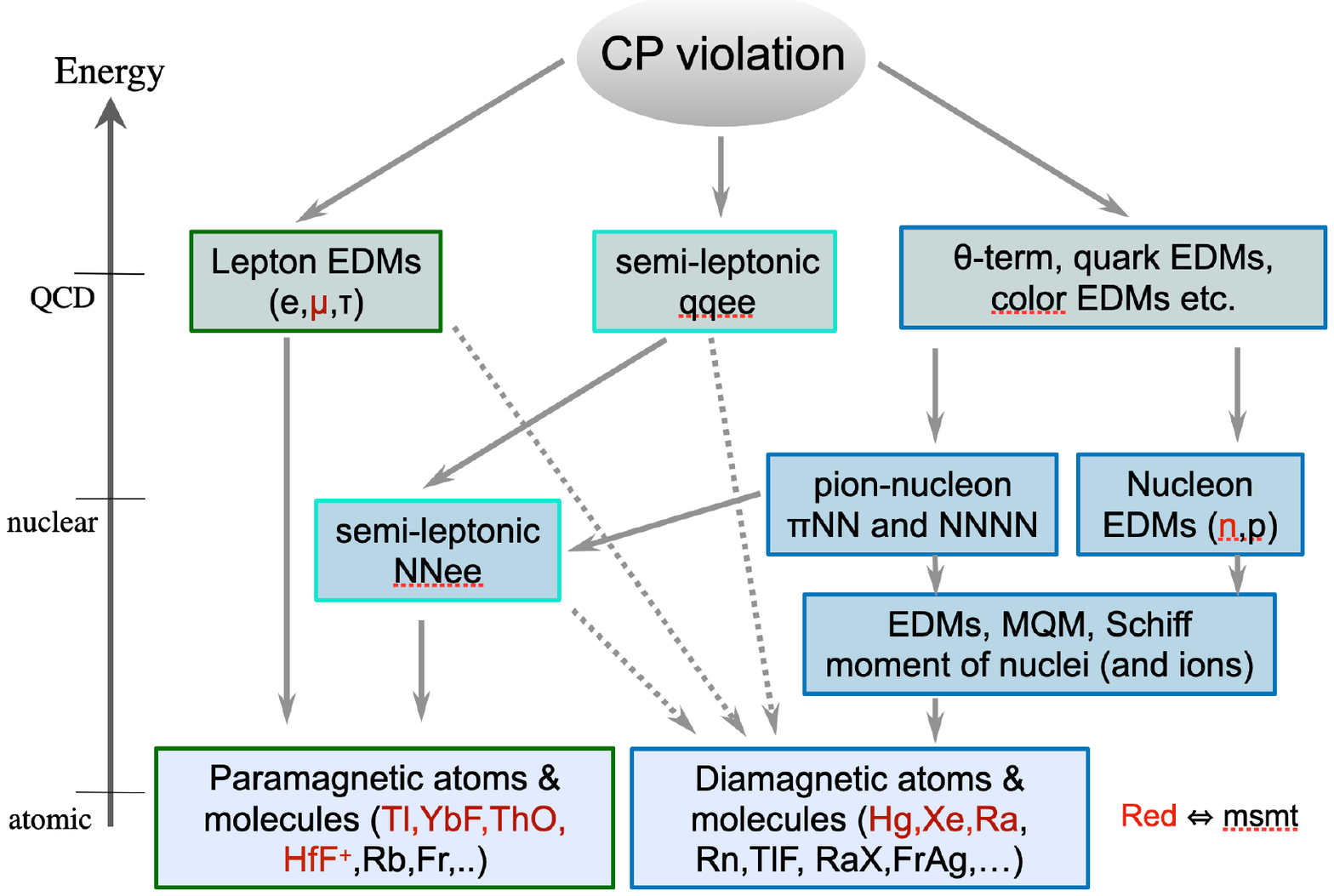}
	\caption{A schematic representation of the hierarchy of effective field theory operators that link fundamental sources of $CP$-violation with observable EDMs. The color-coding reflacts the inheritance patterns for leptonic, semi-leptonic and hadronic $CP$-odd sources respectively. Observables highlighted in red have existing direct measurements that impact the network of constraints.}
	\label{fig:sources}
\end{figure}


 The presence of terms in (\ref{leff}) that do not contain the EM field strength explicitly should not cause confusion: internal hadronic, atomic and finally molecular dynamics ensures the eventual coupling of $F_{\mu\nu} $ to spin, provided that $T$ and $P$ symmetries are broken by these operators.
In particular, the EDMs of nucleons and  
nuclei are also sensitive to the gluonic structure of QCD, starting with the pseudoscalar density $G\tilde G$, that due to the non-perturbative and nonabelian effects of QCD cannot be neglected (as one would neglect the EM pseudoscalar density $F\tilde F$ as a total derivative). The $G\tilde G$ operator {\em cannot} be represented by any perturbative vertex \cite{tHooft:1976rip}, but contributes nonperturbatively. Indeed, the coefficient has two contributions, 
\begin{equation}
     \bar{\th}= \th_{\rm QCD} - {\rm Arg Det}M_q,
\end{equation}
where $M_q$ is the quark mass matrix, which shift in compensating ways under $U(1)_A$ due to the axial anomaly. As a result, $\bar\th$ vanishes in the chiral limit. However, given finite quark masses, the fact that it is dimensionless leads to the so-called strong $CP$ problem in understanding its small size.

Setting aside any contribution from the lowest dimension $\th$-term (removed, for example, by axions \cite{Peccei:1977hh,Weinberg:1977ma,Wilczek:1977pj}) the higher-dimension sources in (\ref{leff}) such as the EDMs ($d_i$) of quarks and leptons and the chromo-EDMs ($\tilde{d}_i$) of quarks lead to numerous
constraints on models of new physics due to their contributions to observable EDMs. The purely gluonic term in the second line, the so-called 3-gluon Weinberg operator \cite{Weinberg:1989dx}, may look like a  higher order generalization of the $\theta$-term, but unlike the $\theta$-term it does lead to perturbative $CP$-odd gluon vertices and generates nucleon EDMs that do not vanish in the chiral limit, $m_q\to 0$. 
Finally, the last term in (\ref{leff}) contains a number of $CP$-odd four-fermion interactions. Note that flavor-conserving $CP$-odd four-fermion operators cannot be built from a single chirality and have to descend from operators involving both $\psi_L$ and $\psi_R$. In theories where chirality flips are associated with extra suppression (for example by small Yukawa couplings of light fermions) their contribution may be less significant. 

Below the 1\,GeV scale, the quark-gluon description is no longer valid as QCD interactions become strong and the relevant degrees of freedom are color-neutral light hadrons, such as nucleons and light mesons. Implementing $CP$-violation in the language of the chiral effective Lagrangian, we list the dominant terms below,
\begin{align}
    {\cal L}_{\rm eff, CP}^{\rm \chi} &= -\frac{i}{2} d_n \bar{n} (F\sigma)\gamma_5 n  -\frac{i}{2} d_p \bar{p} (F\sigma)\gamma_5 p+ \bar{g}^{(0)}_{\pi NN} \bar{N} (\tau \cdot \pi) N + \bar{g}^{(1)}_{\pi NN} \bar{N} \pi^0 N \nonumber \\
    &\quad -\frac{G_F}{\sqrt{2}} C_S^{(0)} \bar{e}i \gamma_5 e \bar{N} N -\frac{G_F}{\sqrt{2}} C_S^{(1)} \bar{e}i \gamma_5 e \bar{N} \ta^3 N + {\cal O}(e^2N^2, N^4,\pi^3,\cdots)
    \label{chiralEFT},
\end{align}
where the weak interaction Fermi constant $G_F$ is introduced for convenience. 
These operators include the electric dipole moments of the nucleons $(d_n,d_p)$, the $CP$-odd trilinear interactions of pions and nucleons, and in the second line semileptonic interactions that play an important role in paramagnetic systems. The matching procedure, or re-expression of Wilson coefficients in Eq.~(\ref{chiralEFT}) in terms of the coefficients of Eq.~(\ref{leff}) is by no means a straightforward exercise, as it requires a nonperturbative analysis in QCD. In the remainder of this section, we briefly review the physics of the three main classes of EDMs, along with the relevant steps that tie the observables to the fundamental $CP$-odd input at the level of effective Lagrangian (\ref{leff}). 

As outlined in Section~1, the physical observables can be conveniently separated into three main categories, depending on the 
physical mechanisms via which an EDM can be generated: EDMs of paramagnetic  atoms and molecules; EDMs of diamagnetic 
atoms; and the neutron EDM. The inheritance pattern for these three classes is represented schematically in Fig.~\ref{fig:sources} and, 
while the experimental constraints on the three classes of EDMs differ by several orders of magnitude, 
it is important that the actual sensitivity to the operators in (\ref{leff}) often turns out to be quite comparable. This
is  due to various enhancement  or suppression factors which are relevant in each case, primarily associated with the  
violation of ``Schiff shielding'' \cite{Schiff:1963zz} in neutral atomic and molecular systems, which constitutes the following important observation.  

If we start with an idealized system of point-like nonrelativistic particles with charge, confined by their own electrostatic force to form a neutral bound state, this can be viewed {\em e.g.} as a simplified version of an atom. Then 
the intrinsic EDMs of point-like particles generate no change in energy in response to an applied external electric field ${\bf E}_{\rm ext}$, once the polarization effects are properly taken into account. Indeed, since such an idealized neutral atom does not accelerate, it is clear that the long-time average electric field experienced by each charged constituent is zero, regardless of the value of ${\bf E}_{\rm ext}$ or, in other words, a compensating internal electric field is developed, ${\bf E}_{\rm int}=-{\bf E}_{\rm ext}$ at the location of a particle. This theorem due to Schiff \cite{Schiff:1963zz} carries over from classical to quantum mechanics, and this shielding effect is one of the key factors for choosing suitable systems for EDM searches. Indeed, the assumptions of the Schiff shielding theorem are violated by relativistic and finite size effects, and this can actually lead to various enhancements in sensitivity. Fig.~2 provides a version of Fig.~1 that highlights the characteristic channels through which Schiff shielding or its violation leads to the enhancement or suppression of sensitivity to EDM observables.

\begin{figure}[t]
	\centering
    \includegraphics[angle=270,width=8cm]{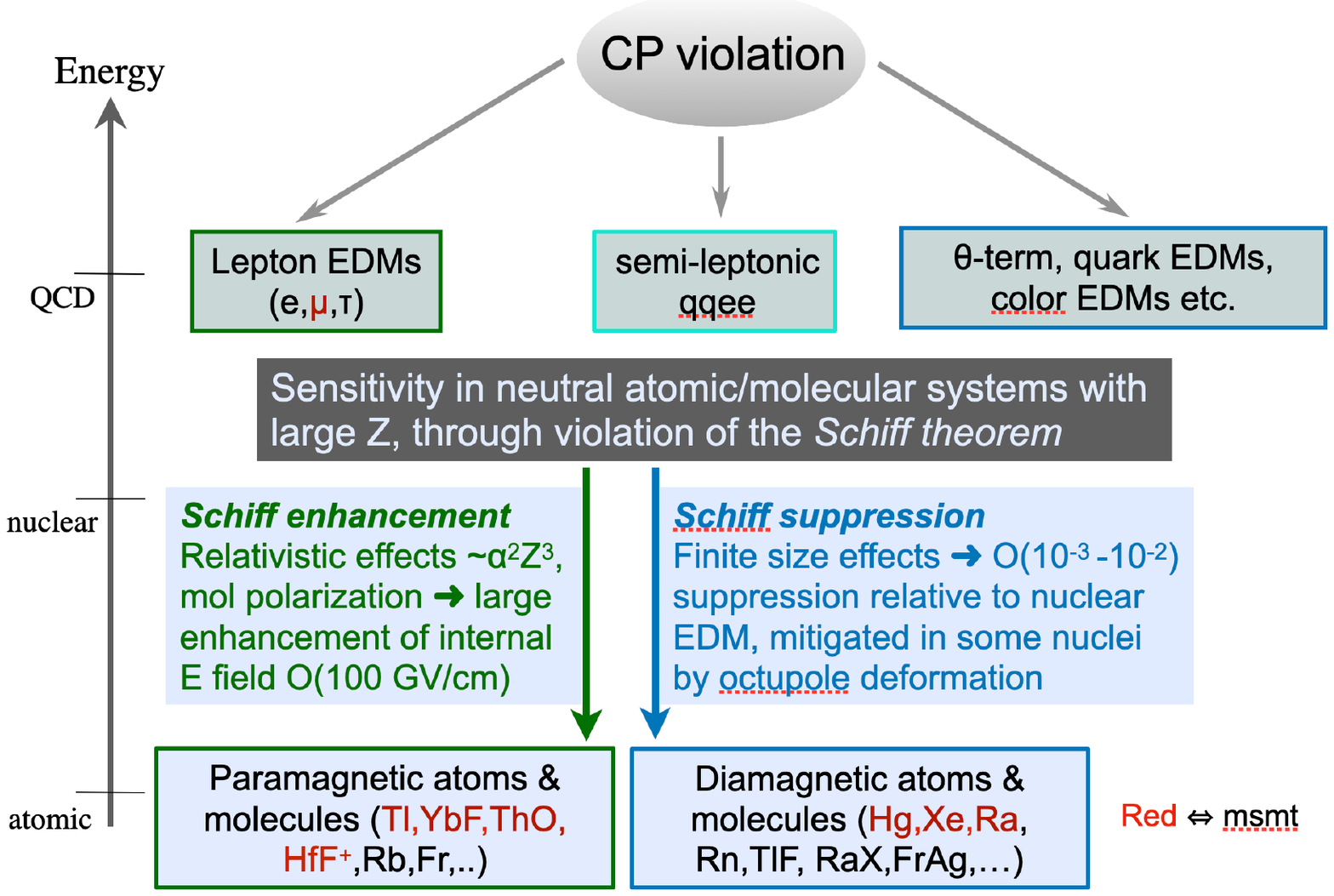}
	\caption{Observable EDMs and the various enhancement and suppression factors that relate measurements to fundamental sources of $CP$-violation.}
	\label{fig:comparison}
\end{figure}

\subsection{EDMs of paramagnetic atoms and molecules}

A class of EDM experiments are performed on neutral atoms and molecules with an uncompensated electron spin, collectively referred to as paramagnetic systems. In certain cases (the ThO molecule, for example) there are an even number of electrons and compensated spins in the ground state, but if EDM experiments are performed on excited states with a nonzero $\langle {\bf S}_e \rangle$, they effectively belong to the same ``paramagnetic" category. 

Paramagnetic atoms and molecules exhibit an important violation of Schiff shielding by relativistic effects associated with the electron's motion. This induces a sensitivity to $CP$-odd interactions of the unpaired electron, namely its intrinsic EDM $d_e$ and also the most relevant $CP$-odd electron-nucleon interaction (\ref{chiralEFT}), namely $C_S^{(0)}\bar e i \gamma_5 e \bar NN$, which in turn is related to the semileptonic four-fermion operators in (\ref{leff}). For large nuclei, the relativistic violation of Schiff shielding can be very large, and indeed the effective internal field seen by the electron can be parametrically larger than the applied field itself. In this sense, Schiff shielding becomes ``Schiff enhancement'' and renders these observables quite valuable as a probe of $CP$-violation. One has parametrically
\cite{Sandars1966,Ginges:2003qt},
\be
 d_{\rm para}(d_e) \sim 10 \al^2 Z^3 (d_e + rC_S^{(0)}), 
\ee
where quantity $r$ is to be discussed shortly. 
For large atoms such as Thallium, the pre-factor represents a large enhancement of the internal electric field (see e.g. \cite{Ginges:2003qt,Khriplovich:1997ga}). 
Thus, the resulting sensitivity to the electron EDM is impressive $d_{\rm Tl} = -585 d_e -   e\ 43 ~{\rm GeV} C_S^{(0)}$, and the Tl EDM was for some time the leading probe of $CP$ violation in the lepton sector.

This relativistic enhancement of the internal field can be boosted even further in polar molecules, where a relatively small applied field can polarize the molecule leading to gigavolt internal fields. Since the response to the applied field is not linear, the induced effect is not an EDM of the molecule itself, and the observable is commonly parametrized via the frequency shift of the relevant levels which scales parametrically as \cite{PhysRevLett.19.1396,Sushkov:1978yj},
\be
 \De \om (d_e) \propto \al^2 Z^3 \frac{M_{\rm mol}}{m_e} d_e,
\ee 
if we just account for the electron EDM as the underlying $CP$-odd source. Starting with the Imperial College group and the polar molecule YbF \cite{Hudson:2011zz}, and with rapid breakthroughts from the Harvard/Yale ACME Collaboration working with ThO \cite{ACME:2013pal,ACME:2018yjb} and most recently the JILA group using the HfF$^+$ ion \cite{Roussy:2022cmp}, the sensitivity to $CP$-odd level splittings has advanced dramatically in recent years.

As for paramagnetic atoms, $CP$-odd electron-nucleon interactions can also provide an important contribution, and so the full frequency shift is conveniently represented in the form,
\be
 \De \om (d_e,C_S) = f_d {\cal E}_{\rm ext} (d_e + r C_S^{(0)}),
\ee
where ${\cal E}_{\rm ext}$ is the external field applied to polarize the molecules, $f_d$ reflects the often very large relativistic violation of the Schiff theorem in polarized molecules (with $f_d {\cal E}_{\rm ext}\sim 23$GV/cm for the JILA experiment using HfF$^+$), and $r$ is determined by the relevant atomic and molecular matrix elements. For the current species of interest, $r$ scales as $r \propto (Z\alpha m_e)G_F$ and is on the order $10^{-20}e$cm. For the most sensitive current experiments (ThO and HfF$^+$) the matrix elements that connect $d_e,C_S$ with $\Delta \omega$ are known quite precisely. It is therefore convenient to interpret single source constraints from paramagnetic systems as limits on either $d_e$ or $d_e^{\rm equiv}$, where
\be
 d_e^{\rm equiv} \equiv d_e + r C_S^{(0)}.
\ee
Since $r$ varies from species to species, at least two precise measurements with different atoms/molecules are required to set independent constraints on both $d_e$ and $C_S^{(0)}$.

\subsection{EDMs of diagmagnetic atoms and molecules}
Diamagnetic atoms have an even number of electrons and  compensated electron spins in their ground states, with the angular momentum of the atom being carried by the nuclear angular momentum. On the experimental side, the overall stability of these systems has enabled long spin coherence times and large statistics, sufficient to push the limits on EDMs of $^{199}$Hg atoms to the record precision for any fundamental EDM \cite{Graner:2016ses}. However, Schiff shielding in this case is violated only by the finite size of the nucleus and differences in the distribution
of charge and the EDM. This is a rather subtle effect that disappears if the nuclear size is taken to zero, resulting in a large suppression of the atomic EDM in relation to the intrinsic EDM of the nucleus: 
\be
 d_{\rm d.\,atom} \sim 10 Z^2 (R_N/R_A)^2 d_{\rm nuc}.
\ee
The suppression by the ratio of nuclear to atomic radii, $R_N/R_A$, generally leads to a suppression of the sensitivity to the
nuclear EDM by a factor of $10^3$ (see e.g. \cite{Ginges:2003qt,Khriplovich:1997ga}). The relevant $CP$-odd nuclear moment in this case is called  the Schiff moment $S$ \cite{Schiff:1963zz}, measured in  $e\,{\rm cm}^3$.
To provide a more complete estimate, we note that atomic calculations determine the EDM as a function of the nuclear Schiff moment, in the form
\be
 d_{\rm Hg} (S) \sim - 10^{-4} S {\rm fm}^{-2}.
\ee
A simple estimate for the Schiff moment $S$ follows from the mean-field potential induced by the $CP$-odd pion-nucleon couplings in the limit where the pion-induced potential is treated as short-range, $\frac{e^{-m_\pi r}}{4\pi r} \rightarrow \frac{1}{m_\pi^2}\delta^3(\vec{r})$. The $T$-odd one-body potential then follows the nuclear charge distribution, and one obtains the following order-of-magnitude estimate \cite{Ginges:2003qt,Khriplovich:1997ga,Engel:2025uci},
\be
 S(\bar{g}_{\pi NN}) \sim 10^{-2} A^{2/3} \frac{eg}{2}\left( \frac{N-Z}{A} \bar{g}^{(0)} - \bar{g}^{(1)}\right) \, {\rm fm}^3,
\ee
where $g\sim 13.3$ is the $CP$-even pion-nucleon coupling. Here we have dropped the isospin tensor pion-nucleon coupling and the dependence on angular momentum quantum numbers for simplicity. Further contributions arise from the intrinsic nucleon EDMs \cite{Dmitriev:2003sc}. Suffice to say that determination of $S$ from the parameters of (\ref{chiralEFT}), particularly for nuclei such as $^{199}$Hg, is among the most challenging topics in nuclear physics. Over the years, the results for $S_{\rm Hg}(\bar g)$ have shown large variations depending on the nuclear physics approach taken for this calculation (see \cite{Engel:2025uci} and references therein).  

Combining atomic $d_{\rm Hg}( S)$, nuclear $S(\bar g_{\pi NN})$, and QCD
$\bar{g}_{\pi NN}^{(1)}(\tilde{d}_q)$, components of the calculation \cite{Khriplovich:1997ga,Pospelov:2005pr,Pospelov:2001ys}, we have
\be
d_{\rm Hg} \sim 10^{-2}\,e\,(\tilde d_u - \tilde d_d)  + {\mathcal O}(d_e,d_q,C_S,C_{qq}).
\label{Hgmaster}
\ee
The overall uncertainty remains rather large, a factor of 2-3 at least, due to uncertainties in the precision of the
nuclear calculation of $S(\bar g_{\pi NN})$ \cite{Engel:2025uci}, and cancellations between 
various contributions at the QCD level.  Nonetheless, a valuable feature of $d_{\rm Hg}$ given the remarkable experimental
precision is its sensitivity to the intrinsic proton EDM and the triplet combination of color EDM operators $\tilde d_q$.

Certain nuclear species exhibit more substantial finite size violations of Schiff shielding, due to so-called octupole deformations \cite{Auerbach:1996zd}. This also has benefits for the calculation of the relevant Schiff moments. A prominent example is $^{225}$Ra, and although the overall sensitivity of current Ra EDM measurements \cite{parker2015measurementatomicelectricdipole} is lower than for $^{199}{\rm Hg}$ due to the radioactive nature of octupole deformed species, this approach shows promise for enhanced sensitivity in the future.

\subsection{Nucleon EDMs}
Direct measurement of the neutron EDM bypasses the nontrivial atomic and molecular enhancement/suppression factors associated with Schiff shielding, and thus provides more direct sensitivity to the underlying sources of $CP$-violation in the QCD sector. This is because the charged constituents of the neutron, {\em i.e.} quarks, are held together not by the electrostatic force but by QCD/strong interactions. Indeed, the earliest EDM measurements by Purcell and Ramsey focussed on the neutron as a probe of parity violation in the strong interactions.

Using chiral symmetries, and taking the Wilson coefficients at 1 GeV, we can write down a simple scaling relation for the dependence on the underlying sources as follows,
\be
 d_n(\bar\th,d_q,\tilde{d}_q) \sim {\cal O}(1)\frac{em_*}{m_n^2} \bar\theta + {\cal O}(1)d_q + {\cal O}(1) e \tilde{d}_q,
 \label{dn}
\ee
where $m_* = \left(\sum m_q^{-1}\right)^{-1} $, and the sum is  extended over all quark masses. In practice,  $m_* \simeq m_u m_d/(m_u+m_d)$. The dependence on $\bar\th$, when combined with the experimental bound on $d_n$, leads to the strong $CP$ problem as it requires a fine-tuning of $\bar\th < 10^{-10}$. We note that the introduction of an axion field via $\bar\th \rightarrow \bar\th + a/f_a$ in the presence of additional $CP$-odd sources does not necessarily relax $\bar\th$ to zero as the axion potential may contain linear terms, leading to $\langle a\rangle/f_a \equiv \th_{\rm ind} \propto \tilde{d}_q$ for example. We will discuss this further below and account for the shift when assuming an axion resolution of the strong $CP$ problem.

Notice that an order of magnitude estimate does not determine the signs of individual contributions to $d_n$, and therefore has limited utility if several sources of $CP$ violation are involved. Therefore, an improvement over (\ref{dn}) is desirable. 
The scaling relations in (\ref{dn}) can be made more precise using a variety of techniques. The dependence on the quark EDMs is equivalent to knowing the corresponding nucleon tensor charges $g_T$,
\be
 d_n(d_q) = g_T^dd_d + g_T^u d_u.
\ee
which have now been computed with sufficient precision using lattice QCD yielding  $g_T^d \simeq (0.78-0.85)$ and $g_T^u\simeq -(0.2-0.25)$ \cite{Gupta:2018lvp,Alexandrou:2019brg}, with normalization scales for $d_{u,d}$ at $2\,$GeV. (The ranges span the central values in \cite{Gupta:2018lvp}, \cite{Alexandrou:2019brg}, while the quoted lattice errors are $O(5-10\%)$.) On accounting for the higher scale normalization at 2 GeV \cite{Gupta:2018lvp}, these results are numerically consistent with earlier estimates using the naive quark model ($g_T^d =4/3$ and $g_T^u =-1/3$) and QCD sum rules. 

Such precision is not yet available for the dependence on other operators, but QCD sum rules techniques \cite{Pospelov:2000bw,Pospelov:2005pr} lead to results that are consistent with the scaling relations above. Estimates obtained using chiral perturbation theory (ChPT) \cite{Crewther:1979pi,Khatsimovsky:1987bb,Dekens:2014jka} are also effective if the answer is enhanced in the infrared by $\log(m_\pi)$. Note that if we account for Peccei-Quinn  relaxation of the axion, where the presence of chromo-EDM sources leads to a nonzero value of $\th_{\rm ind}$ as outlined above (and discussed further in Section~4.1), the contribution of sea-quarks is also suppressed at leading order:
\ba
 d_{n}(\tilde{d}_q,w) &=& (1.1 \pm 0.5)e(\tilde d_d (1\,{\rm GeV}) + 0.5\tilde d_u(1\,{\rm GeV})) + (20\pm 10)\,{\rm MeV}\times e~ w(1\,{\rm GeV}) +{\mathcal O}(C_{qq}).
\label{dn1}
\ea
Note that these results utilize a proportionality to $d_q\langle \bar qq\rangle \sim  m_q\langle \bar qq\rangle \sim f_\pi^2m_\pi^2$, which limits sensitivity to the poorly known light quark masses for a large class of models where $d_q$ scales as $m_q$. 

We conclude this subsection by highlighting that a number of groups are continuing to develop lattice QCD tools to compute the dependence of $d_{n,p}$ and $\bar{g}_{\pi NN}$ on $\bar\th$ and chromo-EDM sources, so there are prospects that the precision of these calculations may improve in the future.

\section{EDMs in the Standard Model} 
\label{Sec3}

As outlined in Section~1, the SM contains two major sources of $CP$-violation. One, a well-established source, namely the complex CKM phase in the vertex of quarks and weak charged currents, is known to give very small EDMs. The other source, the $\theta$-term of the QCD, is capable of inducing very large EDMs that have not been seen. In this section, we review the EDMs resulting for both sources. 

\subsection{EDMs induced by the CKM phase}

In order to serve as an efficient probe of BSM $CP$ violation, one has to make sure that the $CP$-violating source already present in the SM, the CKM phase in the quark sector, generates either very small (or alternatively, calculable) EDMs. Since we deal with a strongly-interacting system of quarks, as the main mediator of $CP$-violation, precision calculations of $d_i(\delta_{\rm CKM})$ are challenging. It turns out, however, that the expected results are about five-to-six orders of magnitude smaller than the current level of experimental precision, leaving an ample window for improvement in experimental precision in the CKM background-free regime. 

We now review the CKM contributions to EDMs. The success of the ongoing flavor physics program has resulted in a 
 rather accurate determination of all CKM parameters, including the $CP$-violating Jarlskog invariant ${\cal J}$ to $O(5\%)$ accuracy,
\be
{\cal J}={\rm Im}[V_{us} V_{cd} V_{cs}^* V_{ub}^*] \simeq 3.1 \times 10^{-5},
\ee
where $V_{ff'}$ are the CKM mixing matrix elements. 
However, predictions for the induced EDMs are complicated by the high perturbative order at which they appear, and by non-perturbative physics at the QCD scale that goes beyond the relatively simple matrix elements discussed in the previous section. 

\begin{figure}[t]
	\centering
	\includegraphics[viewport = 150 150 440 650,angle=0,width=4cm]{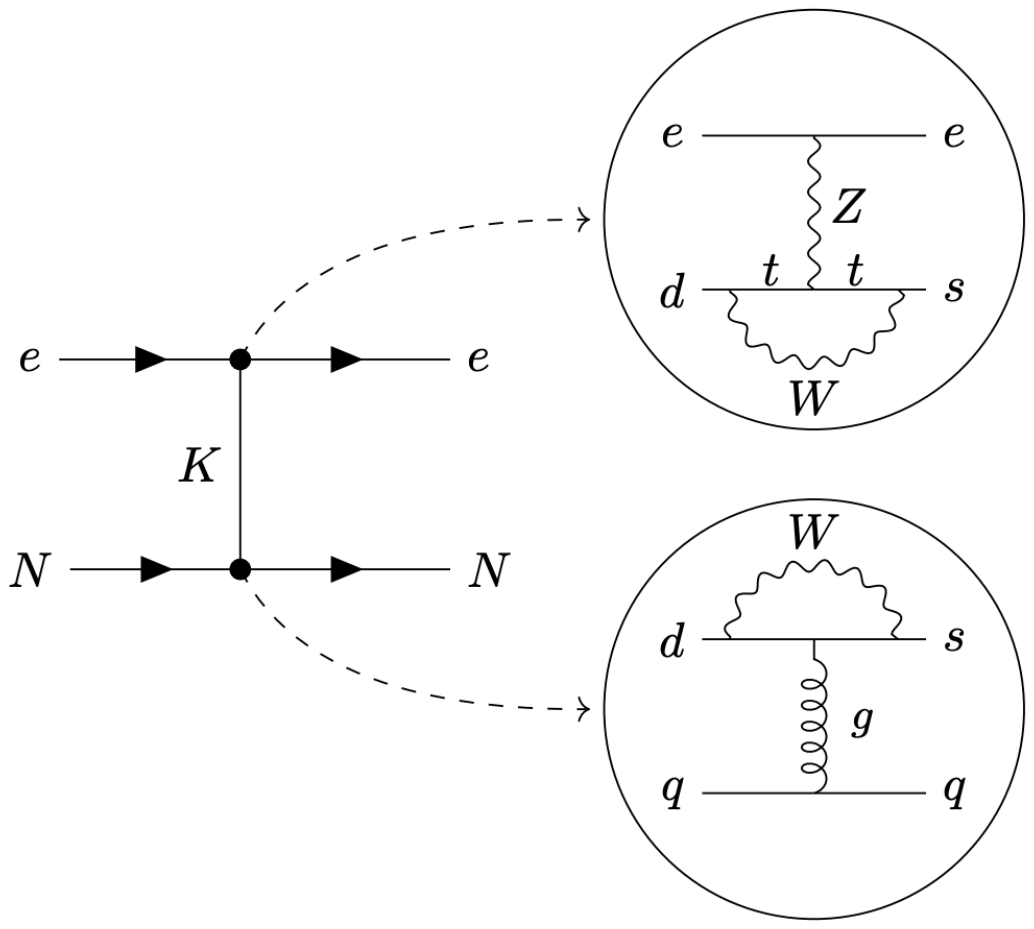}
	\caption{Contribution of the SM CKM phase to the semileptonic Wilson coefficient $C_S$. On the left is the hadronic representation of $O(m_s^{-1})$ diagrams leading to $C_S$, and on the right the effective vertices are resolved as penguin diagrams. The top circle corresponds to an electroweak penguin and scales as $O(G_F^2m_t^2)$. Figure reproduced from \cite{Ema:2022yra}.}
	\label{fig:SM}
\end{figure}

We can again organize the discussion around a simple counting scheme, using the symmetries to estimate the largest viable
contribution to different classes of EDMs. In particular,
CKM contributions to flavor-diagonal observables necessarily vanish at first order in the weak interaction, as any combination of CKM matrix elements in this case is necessarily real, proportional to $|V_{ij}|^2$. Non-vanishing EDMs require all three generations to be involved, which is possible with a minimum of four weak charged current vertices, or equivalently, at order ${\cal O}(G_F^2)$ or EW$^2$.
The EDMs of fundamental fermions also break chiral symmetry, and so must be proportional to a chirality-breaking parameter. Moreover, the antisymmetric flavor structure of ${\cal J}$ leads to further suppression by quark masses of the second generation, as $CP$ violation disappears in the limit of {\em e.g.} $m_c \to m_u$. 
In addition, for reasons related to this antisymmetrization, the lowest order ${\rm EW}^2$ two-loop contributions to quark EDM Wilson coefficients vanish as well \cite{Shabalin:1978rs}. It is worth mentioning that higher order operators (the effective radius of light quark EDMs, Weinberg operator, etc) do not vanish in the lowest EW order \cite{Shabalin:1978rs,Pospelov:1994uf}. 

Quark EDMs therefore require an additional gluon loop, leading to EW$^2$QCD$^{1}$ three-loop order overall that was first reliably calculated by Khriplovich \cite{Khriplovich:1985jr}. Simple estimates suggest \cite{Pospelov:2013sca},
\be
 d_d({\cal J}) \sim e {\cal J} \frac{\al \al_W^2}{(4\pi)^3} \frac{m_d}{m_W^2} \frac{m_c^2}{m_W^2} \sim 10^{-34} e{\rm cm}.
\ee
The appearance of the square of the charm quark mass in the numerator is very common for flavor physics \cite{Glashow:1970gm}. 
Beyond this estimate, more careful calculations have been performed in the literature over the years, predicting $d_d \sim -0.7\times 10^{-34} e{\rm cm}\times (m_d/10\,{\rm MeV}) $ \cite{Khriplovich:1985jr,Czarnecki:1997bu}. 

In considering the intrinsic EDM of nucleons, it is notable that the  charm quark mass is very close to the hadronic mass scale $m_{\rm had}\sim 4\pi F_\pi\sim O(1\,{\rm GeV})$. It is therefore reasonable to ask whether the combination of two flavor-changing transitions with intermediate hadronic states of {\em e.g.} nonzero stangeness may lead to an overall enhancement of EDM amplitudes. Indeed, powerful QCD enhancement mechanisms are at play that amplify $\Delta$(Isospin)=$1/2$ amplitudes \cite{Shifman:1975tn}, so the combination of two $\Delta S =+1$ and $\Delta S=-1$ transitions dominates over the intrinsic quark EDMs in the calculations of the $d_n$ observable  \cite{Gavela:1981sk,Khriplovich:1981ca}.
We can again provide a scaling estimate, based on the underlying symmetries. For the neutron, we find \cite{Pospelov:2013sca}
\be
 d_n ({\cal J}) \sim e {\cal J} \frac{\al_s\al_W^2}{4\pi}\frac{m_{\rm had}^3}{m_W^2} < 10^{-31} e{\rm cm}.
\ee
More recent hadronic model/ChPT reanalysis \cite{Seng:2014lea} is fairly consistent with this estimate, finding $d_n$, with a factor of a few uncertainty, to be of order $10^{-32} e{\rm cm}$.

Lepton EDMs induced by the CKM phase are further suppressed, as they require a closed quarks loop. Moreover, three-loop EW$^3$ diagrams for $d_e$ vanish \cite{Pospelov:1991zt}, and nonzero contributions start at 4-loop order, scaling either as ${\cal O}(\al_W^3\al_s)$ or as ${\cal O}(\al_W^2\al^3)$ \cite{Pospelov:2013sca}. 
Both are similar numerically, with the former scaling as \cite{Pospelov:2013sca},
\be
d_e({\cal J}) \sim e{\cal J} \frac{\al_W^3\al_s}{(4\pi)^4}\frac{m_e m_c^2m_s^2}{m_W^6} < 10^{-44} e{\rm cm}.
\ee
Similarly to $d_n$, the long-distance contributions to $d_e$, arising at ${\cal O}(\al_W^2\al^3)$ are found to be dominant \cite{Yamaguchi:2020eub} bringing $d_e$ to $10^{-40}e{\rm cm}$, with considerable uncertainty. In practice, the largest CKM contribution to paramagnetic EDMs comes not from $d_e$, but from $C_S({\cal J})$. Indeed, as pointed out in \cite{Pospelov:2013sca}, two-photon exchange at EW$^2$EM$^2$ order combines two $\Delta S =\pm 1$ amplitudes and generates an effective electron-nucleon interaction $C_S({\cal J})$ at a level corresponding to $d_e^{\rm equiv}\sim 10^{-38} e{\rm cm}$.

Finally, it has recently been found \cite{Ema:2022yra} that $C_S({\cal J})$ induced at EW$^3$ order represents the dominant contribution to the paramagnetic EDMs, that can be reliably calculated using the flavor $SU(3)$ symmetry and chiral perturbation theory. Fig.\,\ref{fig:SM} shows, schematically, the essence of this calculation. The so-called electroweak penguin diagram generates the $CP$-odd $\Delta S = 1 $ four-fermion operator, that in chiral perturbation theory sources the $(\bar e i\gamma_5 e) K_{S}$ operator. The large size of the top quark mass erases one power of the weak constant in the $G_Fm_t^2$ combination, as is widely appreciated in the flavor literature. Another weak vertex, the $K_S \bar NN $ effective coupling, can be obtained from $s$-wave nonleptonic hyperon decay amplitudes upon the use of flavor $SU(3)$ symmetry. Finally, kaon exchange is enhanced by an $m_s^{-1}$ factor at leading order, and can be systematically calculated/improved using ChPT. In this manner, one arrives at the estimate $G_F C_S \sim {\cal J} \frac{\al_W^3}{(4\pi)^3} \frac{m_e m_t^2}{m_W^6}\frac{\La_{\rm had}^2}{m_s}$, leading to an equivalent electron EDM sensitivity of
 \be 
  d_e^{\rm equiv}({\cal J}) \sim r C_S({\cal J}) \sim r {\cal J} \frac{\al_W^2}{(4\pi)^2} \frac{m_e m_t^2}{m_W^4}\frac{m_{\rm had}^2}{m_s}
  \sim 10^{-35} e{\rm cm}.
\ee
This result shows that SM predictions for EDMs of paramagnetic atoms are dominated by $C_S({\cal J})$. The precision of this prediction, due to the semileptonic nature of $C_S$ is much higher than the corresponding result for $d_n$. 

Finally, for diamagnetic EDMs dominated by the Schiff moment, atomic and nuclear calculations provide a suppression of order $d_{\rm Hg}({\cal J}) \sim 10^{-25} \eta_{np}({\cal J})\, e{\rm cm}$ in terms of the 4-nucleon operator ${\cal L} = \frac{1}{\sqrt{2}} G_F \eta_{np} \bar{N} N {\bar N} i\gamma_5 N$. Accordingly, the atomic EDM scales as follows \cite{Pospelov:2013sca},
\be
 d_{\rm Hg} ({\cal J}) \sim {\cal J} G_F m_{\rm had}^2 \times 10^{-25} e{\rm cm} < 10^{-35} e{\rm cm},
\ee
where for $\eta_{np}$ one can use again the combination of strangeness-changing transitions \cite{He:1992jh}. 

All the EDM estimates discussed in this section, induced by the SM CKM phase, are well below current experimental
sensitivity. The evolution of experimental sensitivity in recent years relative to the CKM background is shown for comparison in Fig.~\ref{fig:progress}. One may conclude that progress of roughly five orders of magnitude in experimental precision is required before the quark-sector CKM phase becomes a non-negligible background.

\begin{figure}[t]
	\centering
	\includegraphics[angle=270,width=11cm]{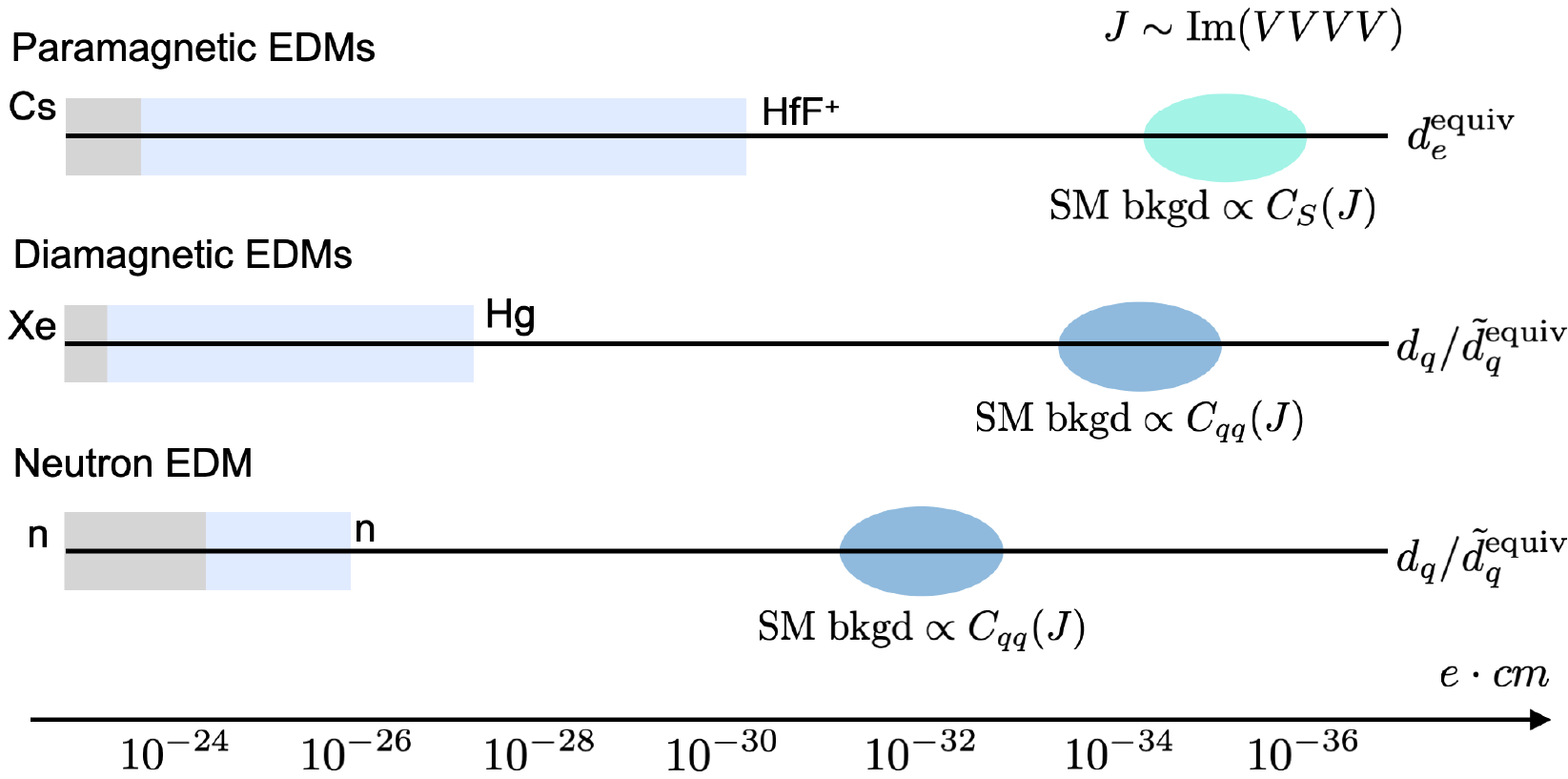}
    \vspace*{-0.5cm}
	\caption{Evolution of precision EDM measurements, from the limits (in gray) circa 1985 through to today (in blue). The SM CKM background contributions are shown for comparison.}
	\label{fig:progress}
\end{figure}

\subsection{EDMs induced by the QCD $\theta$ angle  }

The observable nature of $\bar\theta$ is directly related to the solution of the $U(1)_A$ problem of QCD \cite{Shifman:1979if}. In the chiral approach, where $m_{u,d,s}\ll m_{\rm hadr}$, one should find a nonet of Nambu-Goldstone bosons with $m_{\rm NGB}^2\propto m_q$ that are massless in the chiral $m_q \rightarrow 0$ limit. In practice, the mass of the flavor singlet boson, the $\eta'$ meson, deviates from this prediction, manifesting a non-perturbative feature of QCD. Consequently, the suppression of $\eta'$-mediated interactions renders observable quantities $\bar\theta$-dependent. 

This central feature of all $\bar\theta$-dependent phenomenology (vacuum energy and axion mass, $CP$-odd $\pi NN$ couplings, and EDMs) can be demonstrated explicitly, taking the two-flavor case for simplicity. An important prerequisite for this analysis is the analyticity of QCD at $\bar\th=0$, and indeed evidence for this has accumulated from a variety of channels since the 1970's.\footnote{In contrast, there is evidence of the non-analyticity of theories such as QCD at $\bar\th=\pi$, where $CP$ may be spontaneously broken (see e.g. \cite{Gaiotto:2017tne}).} Consequently, despite the more complex implications of $\theta$ in determining quantum mechanical $\th$-vacua, the physical quantity $\bar\th$ can be analyzed in the same manner as other parameters which vanish in the chiral limit. The light quark mass sector of QCD, upon use of an anomalous chiral rotation to remove the $\theta G\tilde G$ operator, takes the form: 
\begin{eqnarray}
    {\cal L}_{m} = -m_u\bar uu -m_d\bar dd -m_*\bar \theta(  \bar ui\gamma_5u + di\gamma_5d)+\frac12\bar\theta^2m_* \left(\frac{\bar uu + \bar dd}{2}\right)+\frac12\bar\theta^2m^2_* \frac{m_u-m_d}{m_u+m_d}\left(\frac{\bar uu - \bar dd}{2}\right)+{\cal O}(\bar\theta^3).
\end{eqnarray}
This expression retains terms up to quadratic order in $\bar\th$, and the entire ${\cal L}_m$ can be thought of as a small perturbation due to the smallness of the quark masses. Chiral symmetry breaking generates nonzero values for quark scalar densities, $\langle 0|\bar uu|0\rangle = \langle 0|\bar dd|0\rangle  \equiv  - |\langle \bar qq\rangle |$, and these condensates directly contribute to the vacuum energy at order $O(m_*)$. However, the flavor singlet ``would-be" Goldstone boson can be created from the vacuum by the operator $m_* (\bar ui\gamma_5u + \bar di\gamma_5d)$, and {\em if} the propagator of this Goldstone mode at zero momentum is indeed proportional to $m_q^{-1}$, then the pole diagram is also of $O(m_*)$, canceling the contact contribution. Nonperturbative effects (which may be realized as instantons in similar theories with a controlled semi-classical regime) \cite{Shifman:1979if}) keep the mass of the singlet boson finite in the strict chiral limit, $m_\eta^2 \to m_0^2$ at $m_q\to 0$, and in terms of this parameter, the $\theta$-dependent part of the vacuum energy density is given by
\begin{eqnarray}
    E_{\rm vac}(\bar \theta) = \frac12\bar\theta^2m_* |\langle \bar qq\rangle |\times\left( 1-\frac{4m_*B}{m_0^2+4m_*B}\right),
    \label{Evac}
\end{eqnarray}
where $B= F_\pi^{-2}|\langle \bar qq\rangle| = m_\pi^2(m_u+m_d)^{-1}$. If the $U(1)_A$ symmetry is restored, {\em i.e.} $m_0\to 0$, then the $\theta$-dependence of the vacuum energy disappears \cite{Shifman:1979if}. In the opposite, and physical, limit of $m_* B\ll m_0^2$, the second term in the parentheses of Eq.~(\ref{Evac}) can be neglected, yielding the conventional nonzero result. 

\begin{figure}[t]
	\centering
	\includegraphics[scale=0.5]{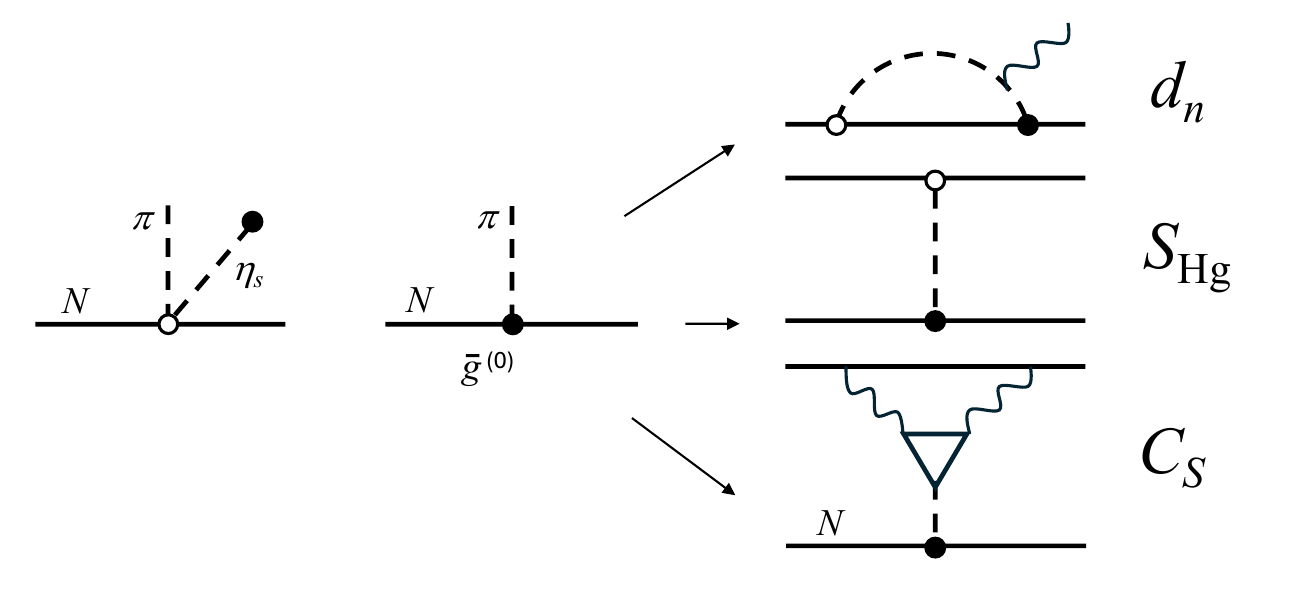}
	\caption{The $\bar\th$ parameter of QCD induces a $CP$-odd vertex between a pion and the nucleon represented by a black dot in the second diagram from the left. If the $U(1)_A$ symmetry is restored, then the first (singlet $\eta_s$ meson re-scattering) diagram exactly cancels the second. In the real world with $U(1)_A$ violated nonperturbatively, the second diagram dominates, and contributes to all three classes of observable EDMs, represented schamatically on the right. White dots and the triangle denote the conventional coupling of pions to nucleons and photons.}
	\label{fig:ThetaEDMs}
\end{figure}

The logic can be applied to the calculation of the $CP$-violating  $\pi \bar N N$ coupling constant \cite{Pospelov:1999mv}. In Fig.\,\ref{fig:ThetaEDMs}, we see that in the combination of the first two diagrams, which would sum up to zero should $U(1)_A$ be restored, the second one is dominant in real-world QCD leading to the prediction of the $CP$-odd vertex ${\cal L}_{{\rm eff},CP} = \bar{g}^{(0)}_{\pi NN} \pi^a\bar{N} \tau^a N$:
\begin{eqnarray}
    \bar{g}^{(0)} \simeq -\frac{m_*\bar\theta}{F_\pi}\times \langle p| \bar uu -\bar dd | p\rangle \simeq - 0.017 \bar \theta,
\end{eqnarray}
where matrix element can be extracted from lattice QCD calculations \cite{BMW:2014pzb} or from phenomenological evaluations based on the flavor $SU(3)$ pattern of baryon mass splitting \cite{Crewther:1979pi}, and we used $m_u/m_d \simeq 0.48$. This vertex can source all three categories of EDMs: the ChPT loop of charged pions induces $d_n$; tree-level pion exchange results in the Schiff moment $S_{\rm Hg}$; and finally an effective two-loop $\pi^0$-electron coupling induces $C_S^{(1)}$. Note that the result for $d_n(\bar\theta)$ appears to be relatively robust: the ChPT result \cite{Crewther:1979pi} was tested using an independent QCD sum rule calculation, yielding similar numerical results \cite{Pospelov:1999ha,Pospelov:1999mv,Ema:2024vfn}, and agreeing with an earlier order-of-magnitude estimate \cite{Shifman:1979if} in Eq. (\ref{dn}). Efforts to rigorously compute $d_n(\bar\th)$ using lattice QCD techniques (see e.g. \cite{Liu:2024kqy}) are underway which promise to further test this result and improve its precision.

Summarizing the predictions and current constraints from the three groups of EDMs discussed above,
\begin{eqnarray}
    d_n \simeq 2\times 10^{-16}\,\bar\theta\,e{\rm cm}~~&\rightarrow&~~|\bar\theta| < 1\times 10^{-10}~~~\cite{Abel:2020pzs}\\
    S_{\rm Hg} \simeq 2\times 10^{-3} \,\bar\theta \,e{\rm fm}^3~~&\rightarrow&~~|\bar\theta|< 1.5\times 10^{-10}~~~\cite{Graner:2016ses}\\
    C_S \simeq 0.03 \,\bar\theta ~~&\rightarrow&~~|\bar\theta| < 2\times 10^{-8}~~~\cite{Roussy:2022cmp}
    \label{cstheta}
\end{eqnarray}
All the above results imply that $\bar\th$ is necessarily small without providing any reason why it is so. Note that despite paramagnetic EDMs currently providing a limit that is $O(100)$ less stringent than $d_n$ \cite{Flambaum:2019ejc,Mulder:2025esr}\footnote{The number in Eq.\,\ref{cstheta} has been slightly updated compared to \cite{Flambaum:2019ejc} due to subsequent strengthening of the experimental limit \cite{Roussy:2022cmp}.}, the rapid improvement of this category of measurements \cite{ACME:2018yjb,Roussy:2022cmp}, gives reasons to believe that one day they may become competitive to hadronic EDMs in their sensitivity to $\bar \theta$. Similarly to the case of CKM contributions, $C_S$ is dominant in $d^{\rm equiv}_e$ over the proper $d_e(\bar \theta)$ contribution \cite{Ghosh:2017uqq}. The limits from the Hg EDM are more uncertain than the results for $d_n$ and $C_S$. Currently they are based on the $S(\bar g^{1)}_{\pi NN}(\bar \theta))$ estimate \cite{Engel:2013lsa} but should also include the better known contribution from $S(d_n(\bar\theta))$. Barring an accidental cancellation, the limit on $\bar \theta$ from the Hg EDM measurement may be as strong as from $d_n$.

\section{EDM constraints on new physics}
\label{Sec4}

The EDM limits and EFT analysis discussed above lead to a network of constraints on quark, gluon and lepton operators at the GeV scale, as illustrated schematically in Fig.~1. These constraints have improved dramatically with improving experimental precision over the past 30-40 years.  
The precision of EDM calculations still varies from 10-15\% for a variety of atomic and molecular calculations, to ${\cal O}(1)$ or worse for calculations of Schiff moments in nuclei such as Hg and a variety of calculations at the QCD level. 
Nonetheless, setting aside issues of calculational precision, these EDM limits provide an essentially model-independent set of 
constraints on new $CP$-violating physics which may generate these operators at the GeV scale. EDM measurements, even 
without a positive detection, therefore provide a suite of stringent constraints on new sources of $CP$-violation. 

\subsection{EDM implications of the strong $CP$ problem and possible solutions}

Our understanding of $CP$ conservation in the strong interactions rests on the success of ChPT, the understanding of nonperturbative QCD obtained in the large-$N$ limit and the resolution of the $U(1)_A$ problem.\footnote{This understanding, initially developed in the 1970s, can be viewed as a vindication of the original idea of Purcell and Ramsey in 1950: lacking knowledge of the fundamental theory behind the strong interactions, we cannot make a theoretical claim of parity conservation in the same way as can be done for electromagnetism.}. On one hand, a quantum mechanical super-selection rule forbids transitions between vacua with different values of $\theta$ (and thus different vacuum energies). On the other hand, due to the experimental EDM limits, we know that $\bar\theta$ must be very small. This presents a naturalness problem: $\bar \theta = 0 $ is perfectly valid, but why would an arbitrary phase parameter that can be $O(1)$ be below $10^{-10}$? Moreover, since $\bar\th$ is dimensionless, arbitrarily heavy new physics can contribute to $\bar \theta$ without any power-like suppression. 

Significant theoretical effort has been devoted to resolving the strong $CP$ problem, and even more to the physics and phenomenology of the most promising solution based on axions. Here we review a narrow subset of this field, as it relates back to the question of EDM observables. In particular, approaches to resolving the strong $CP$ problem can be subdivided into the three different categories, which we list below; the implications of these solutions for EDMs are then discussed in the remainder of this subsection: 

\begin{enumerate} 

\item One of the light quark masses is zero, {\em e.g. $m_u=0$}. This results in $m_*\to 0$ and all the $\theta$-dependent observables will vanish. 

\item The vacuum angle is promoted to a new dynamical field, $\bar\theta \to a/f_a$. An axion field $a$ will seek its minimum energy, and since min($E_{\rm vac}(\bar \theta)$)=0, over time scales $\propto f_a$, the value of the axion field will satisfy $a/f_a\ll 1$, regardless of the initial position of $a$. In this way, any large angle at early times can be erased.

\item Finally, there have been attempts to consider the strong $CP$ problem purely from the point of view of discrete symmetries. If one demands, for example, that the fundamental UV theory posseses exact $P$ or $CP$ symmetry (see {\em e.g.} \cite{Mohapatra:1978fy,Nelson:1983zb,Barr:1984qx,Hiller:2001qg,Hook:2014cda}), subsequent spontaneous breaking of these symmetries can be arranged in such a way that $\bar\theta$ remains small, and the strong $CP$ problem is under control. At the same time, the CKM phase needs to be generated at the ${\cal O}(1)$ level.

\end{enumerate}

The $m_u=0$ solution (1) is of course the most economical. However, it is at odds with a multitude of ChPT and lattice QCD data suggesting that $m_u/m_d\simeq 0.5$. For example, one of the latest evaluations \cite{Colangelo:2018jxw} puts $m_u/m_d = 0.45(3)$, which is clearly incompatible with $m_u=0$. Notice that for this solution to be valid in theory, $d_u$, $\tilde d_u$, and some of the four-fermion operators in (\ref{leff}) would have to vanish. Otherwise, higher-order dimensional transmutation will allow the regeneration $d_u, \tilde d_u \to m_u$, spoiling the vanishing of $m_*$. Therefore, if the $m_u=0$ scenario for resolving strong $CP$ were to have been realized, nonzero EDMs would be generated only from the lepton and down-quark/gluon sector ($d_e$, $d_d$ etc).

The axion solution (2) to the strong $CP$ problem has enormous implications for particle physics and cosmology. It also impacts EDMs by eliminating all $dim=4$ contributions and restoring the decoupling of heavy states that potentially break $CP$ at very high energies. It is worth mentioning, however, that an {\em induced} $\theta$ term reappears in this case through the linear shift of the axion vacuum in the presence of extra $CP$ violating sources \cite{Bigi:1991rh,Pospelov:1999ha,Okawa:2021fto}: 
\begin{eqnarray}
    \theta_{\rm ind} = - \frac{\int d^4x \langle T(\frac{\alpha_s}{8\pi} G\tilde G(0), {\cal L}^{\rm CP-odd}_{dim \geq 5}(x) \rangle }{\int d^4x \langle T(\frac{\alpha_s}{8\pi} G\tilde G(0), \frac{\alpha_s}{8\pi} G\tilde G(x) \rangle}~~\to~~ - \frac12\left( \sum_{q=u,d,s}\frac{\tilde d_q}{m_q}\right)\frac{\langle0|\bar q (G\sigma)q|0\rangle}{\langle0|\bar q q|0\rangle} + \cdots
\end{eqnarray}
The denominator of the first expression is known as the topological susceptibility of the QCD vacuum, which is directly related to $d^2E_{\rm vac}/d\theta^2$. In the second expression, we present the induced $\theta$ parameter generated by the color EDMs of light quarks, and the ratio of two condensates is estimated to be $\simeq 0.8\,$GeV$^{2}\sim m_{\rm had}^2$. Diagrammatically, this effect arises due to an axion tadpole with ${\cal L}^{\rm CP-odd}_{dim \geq 5}$ mediating a transition between the axion field and the vacuum. Notice that the presence of a nonzero $\theta_{\rm ind}$ does not destroy the decoupling of heavy BSM physics: it vanishes together with the higher dimensional source if the energy scale of $CP$ violation is taken to be very large, {\em i.e.} $\tilde d_q \propto m_q/\Lambda_{CP}^{2}$. 

The presence of axions and $dim\geq 5$ $CP$-violation gives rise to a phenomenon closely intertwined with EDMs. A $CP$-odd non-derivative axion nucleon coupling develops, $\bar g_{aNN} a \bar NN$ \cite{Moody:1984ba}, that can be tested with precision gravity experiments, as the range of the axion force for a phenomenologically viable axion can be macroscopic, {\em e.g.} $\lambda_a  \geq$1\,mm. Since $\bar g_{aNN}$ is also induced by  higher-dimensional operators ({\em e.g.} $\tilde d_q$ etc), it currently appears that EDMs provide more constraining power \cite{Pospelov:1997uv} than experimental tests of the axion-generated fifth force. 

Finally, the symmetry-based approach (3) to solving the $CP$-problem is more controversial, not least because it imposes a UV-condition on the properties of the $\theta$-vacua associated with long-range QCD dynamics. Nonetheless, the assumption that $\bar \theta = 0$ is a legitimate starting point at very high energy scales has interesting consequences for EDMs. The radiative corrections to $\bar\theta$ from the CKM phase are known to be small \cite{Khriplovich:1985jr,Ellis:1978hq}. However additional dynamics associated with the spontaneous breaking of {\em e.g.} $CP$ invariance could generate finite corrections to the $\theta$-term that do not decouple. In other words, one needs a fully engineered model, with control over the field content, possible interactions and especially the absence of additional sources of $CP$ violation to almost arbitrarily high scales. Models that employ a heavy axion like particle, with $m_a$ sourced by something other than the usual quark-gluon dynamics at a GeV, suffer from the same conceptual problem of extreme UV sensitivity. Indeed, the axion minimum can be displaced away from zero to an observable level by $CP$-violating operators generated at very high scales \cite{Bedi:2022qrd,Csaki:2023ziz}. 

We end this subsection by noting that in models with high-scale discrete symmetries, or with explicit UV breaking of the Peccei-Quinn (PQ) symmetry, the $\theta$-term is expected to dictate the manifestation of $CP$ in the quark sector more broadly. This adds extra impetus for more refined calculations of the EDMs as a function of $\bar\theta$. Indeed, in the speculative scenario that nonzero EDMs were to be discovered with multiple detections {\em and} the calculation of relevant observables was brought to a precision of $\sim 10\%$, it is conceivable that a ``$\theta$-pattern" could be recognized among different observables, not unlike the manifestation of $\delta_{\rm CKM}$ effects across the entire range of flavor physics data. 

\subsection{EDMs vs Higgs and electroweak physics}

In the remainder of the this section we will assume that the strong $CP$ problem is resolved by axions for concreteness, so that we can interpret EDM observables as  being sensitive to higher-dimensional $dim\geq 5$ operators in the conventional Wilsonian picture. EDMs then directly probe high-energy scales. 

The discovery of the Higgs boson at the LHC in 2012 was an impressive scientific and technological achievement, and the subsequent measurement of many of its properties with growing precision has cemented the status of the Standard Model. It is therefore natural to parametrize the sensitivity to high-scale new physics in terms of EFT operators normalized at the electroweak scale. Here 
we concentrate on a few representative examples that capture $CP$-violating interactions of new heavy resonances. We start with the $CP$-violating Higgs interaction with photons that descends from the following effective Lagrangian:
\begin{eqnarray}
\label{ch}
    {\cal L}^{CP-odd}_{h\gamma\gamma} = H^\dagger H\times \left(\frac{g_1^2}{e^2\Lambda_{h\gamma}^2} a_h \times B_{\mu\nu}\widetilde B_{\mu\nu} +  b_h \times \frac{g_2^2}{e^2\Lambda_{h\gamma}^2}W^a_{\mu\nu}\widetilde W^a_{\mu\nu} \right)~\to ~ \frac{ c_h v_{\rm EW}}{\Lambda_{h\gamma}^2} \times hF_{\mu\nu}\widetilde F_{\mu\nu},
\end{eqnarray}
where the coefficient $ c_h = a_h + b_h $ could in principle be subsumed into the definition of $\Lambda_{h\gamma}$, $v_{\rm EW} = \sqrt{2}\langle  H\rangle=246$\,GeV is the Higgs vacuum expectation value, and $h$ is the physical, 125\,GeV mass, Higgs boson field. As expected, the Wilson coefficient scales as $\langle H\rangle/\Lambda^2$. An interesting aspect of this effective operator is that it always increases the overall rate of the Higgs decay $h\to 2\gamma$ as the main SM amplitude is $CP$-even, and therefore ${\cal L}^{CP-odd}_{h\gamma\gamma}$ can be constrained from precision Higgs studies. Independently of this constraint, EDMs already impose stringent limits on any $CP$-odd coupling of the Higgs to photons \cite{McKeen:2012av}, through the intermediary of the $hF\tilde{F}$ operator via the left-hand diagram shown in Fig.~\ref{fig:Higgs}. The explicit calculation \cite{McKeen:2012av} of an induced dipole contains an ultraviolet logarithmic divergence indicating the mixing of effective operators. Taking the electron EDM as an example, one finds
\begin{eqnarray}
    d_e \simeq c_h\frac{e m_e}{4\pi^2\Lambda^2_{h\gamma}}\log\left(\frac{\Lambda_{\rm UV}^2}{m_h^2}\right)~\to~ \frac{ c_h}{\Lambda^2_{h\gamma}} < \frac{1}{({\rm 800\,TeV})^2},
    \label{hgam}
\end{eqnarray}
where the logarithmic divergence is cut off at $\Lambda_{\rm UV} \sim 20$\,TeV. 
The sensitivity scale of 800\,TeV looks very impressive, but any UV completion of (\ref{ch}) would involve at least one loop factor, suggesting that $c_h < \alpha_W/(4\pi)$ and implying that the maximal scale probed by such an operator is $\sim 40$\,TeV. 

The heaviest particle in the SM, the top quark, may allow for larger $CP$-violating couplings. Of particular interest is the $CP$-odd top-Higgs coupling, $h(\bar t i\gamma_5 t)$, that follows from the following effective operator, 
\begin{eqnarray}
    {\cal L}_{htQ} = -H^\dagger H\times \frac{y_t\exp{(i\phi_t)}}{\Lambda^2_{ht}} \times \overline{t}_R {Q}_{3L}H+(h.c)~\rightarrow ~ 
    {\cal L}_{htt}^{CP-odd} = \sin(\phi_t)\frac{ m_t v_{\rm EW}}{\Lambda^2_{ht}}\times h (\bar t i\gamma_5 t),
    \label{ht}
\end{eqnarray}
where the top quark Yukawa coupling $y_t$ is introduced for convenience, and $Q_{3L} $ is the left-handed doublet of the third generation. In deriving this interaction vertex, we assumed that the $dim=4$ top quark Yukawa coupling is real. Notice that the presence of $H^\dagger H/\Lambda^2$ in (\ref{ht}) is crucial for the existence of the $CP$-odd vertex, as otherwise the complex phase of $\overline{t}_R {Q}_{3L}H$ can be absorbed into the definition of the top quark Yukawa coupling. 

Integrating out the heavy states, $t$ and $h$ in this particular case, leads to two-loop induced EDMs and color EDMs of light fermions. The loop functions of these so-called Barr-Zee diagrams \cite{Barr:1990vd} are well known. If the top quark can be considered parametrically heavy, the integration of the top loop leads to effective interactions in Eq.~(\ref{ch}). Taking, for example, the EDM of the electron induced at two loops via $ht\gamma $ mediation, one finds (see {\em e.g.} \cite{Huber:2006ri}),
\begin{eqnarray}
    d_e^{\,ht \gamma } =\sin(\phi_t)\times \frac{em_e}{\Lambda^2_{ht}} \times \frac{\alpha}{6\pi^3} g(m_t^2/m_h^2)  ~\rightarrow ~ 4.1\times 10^{-30}\,e{\rm cm}\times \left(\frac{\rm 10\,TeV}{\Lambda_{ht}}\right)^2\times \left(\frac{\sin(\phi_t)}{1/\sqrt{2}}\right).
    \label{hgt}
\end{eqnarray}
The loop factor here is $g\simeq 1.4$. 
While this is one among a family of similar two-loop contributions, it illustrates that even with the large numerical suppression of the loop factors, the current EDM sensitivity to $CP$-odd effective operators stretches to the $O(10\,\rm TeV)$ scale. For an up-to-date account of the impact of two-loop Barr-Zee diagrams in the two-Higgs doublet model with $CP$ violation see {\em e.g.} Ref.\,\cite{Altmannshofer:2025nsl}. 

Thus far, our review in this section has remained at the level of effective $CP$-odd operators. Nevertheless, it is also instructive to ask the question of what it takes to UV-complete Eqs.~\ref{ch}) and (\ref{ht}), {\em i.e} to present a renormalizable model that resolves these effective interactions. Indeed, many UV complete examples are known, starting with multi-Higgs doublet models, supersymmetric models, models with an extended gauge sector, etc. One common feature of models that induce (\ref{ch}), (\ref{ht}) and similar operators is that the new physics states that have been integrated out are charged under the SM gauge group(s). In $CP$-even channels, effective electroweak operators can be induced by completely neutral particles. For example, two types of interactions, $A (H^\dagger H) S$ and $\lambda S^3$, will generate $CP$-even higher-dimensional interactions $(H^\dagger H)^3$ and $H^\dagger H\times\overline{t}_R {Q}_{3L}H$ upon integrating out the neutral scalar $S$. In contrast, $CP$-odd operators $\propto \sin(\phi_t)$ in (\ref{ht}) will not be induced. In order to generate $CP$-odd combinations, one needs, for example, a new heavy Higgs doublet $H'$ and  to combine {\em {e.g.}} $(H^\dagger H)(H^\dagger H')$ and $e^{i\phi_t} \overline{t}_R {Q}_{3L}H'$. Therefore, one can draw an important general conclusion: EDMs provide sensitive probes of $CP$ violation in the interaction of the SM with {\em new EW-charged states}.  

$CP$ violation at the electroweak scale also finds direct use within models of {\em electroweak baryogenesis} \cite{Kuzmin:1985mm,Nelson:1991ab,Cohen:1993nk,Morrissey:2012db}. It complements two other necessary ingredients: thermal effects that break the $B+L$ number through sphalerons and non-equilibrium dynamics related to a first order EW phase transition. The first order phase transition generates bubbles inside which the EW vacuum is in the broken phase where the sphaleron processes are highly suppressed. The domain wall separating the two phases can be thought of as a non-trivial Higgs field profile interpolating between $0$ and $v_{\rm EW}$. Inside a moving domain wall, the second operator in Eq. (\ref{ch}) generates a chemical potential for the EW Chern-Simons charge and biases the electroweak sphaleron processes to produce more quarks than antiquarks for the appropriate sign of $CP$ violation. $CP$-odd operators of the type shown in (\ref{ht}) will also lead to an asymmetry between transmission and reflection coefficients of $t$ and $\bar t$, resulting in the accumulation of $t$ quarks inside the broken phase, while in the unbroken phase the excess of anti-quarks is diluted between leptons and quarks by sphaleron processes. 

In the SM with the observed value of the Higgs boson mass, the EW phase transition is instead a smooth cross over from the symmetric to the broken phase, and as a result pure SM EW baryogenesis is not viable and new physics is required to enhance the strength of the transition. Typically this is achieved by extending the SM with light scalar fields, which can be either charged or neutral under the SM gauge group. However, neutral fields cannot serve as an efficient $CP$-odd source, and one needs to extend the model further to include relatively light fields, with mass less than a few hundred GeV, having SM charges and $CP$-violating phases in their interaction with the SM electroweak sector. EDMs are then generated, often at two loop level as in (\ref{hgt}). If the relevant scales associated with these new particles are below 1 TeV, EDMs constrain the relevant $CP$-phases, such as $\phi_t$ in (\ref{ht}, too much to enable efficient baryogenesis \cite{Huber:2006ri}. While this conclusion holds for most of the EW baryogenesis models discussed in the literature, there are some notable exemptions that allow the generation of a sufficient baryon asymmetry while conforming to the EDM constraints. Approaches include carefully placing the $CP$-violating source among particles that have little or no interaction with light fermions, placing $CP$ violation in flavor off-diagonal transitions, and/or through the resonant enhancement of small interactions for heavy on-shell particles that does not occur inside EDM-generating loops. For a small subset of the relevant literature, see {\em e.g. } Refs.~\cite{Kozaczuk:2012xv,Kanemura:2023juv,Aiko:2025tbk}. 
 
\begin{figure}[t]
	\centering
    \vspace{-2cm}
	\includegraphics[angle=0,viewport = 400 -280 1400 600, width=5.5cm]{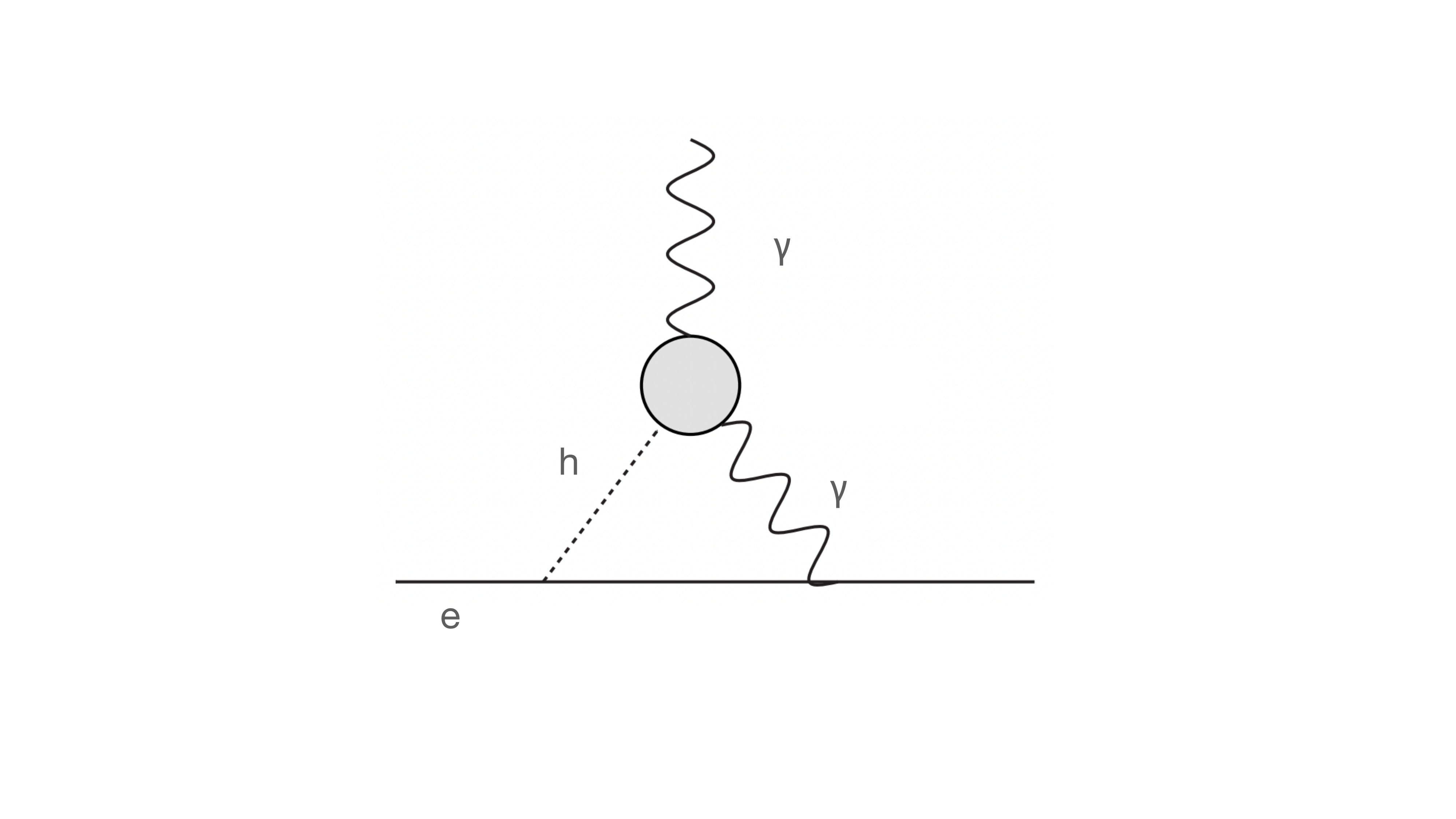}\hspace*{1cm}
    \includegraphics[angle=0,width=7cm]{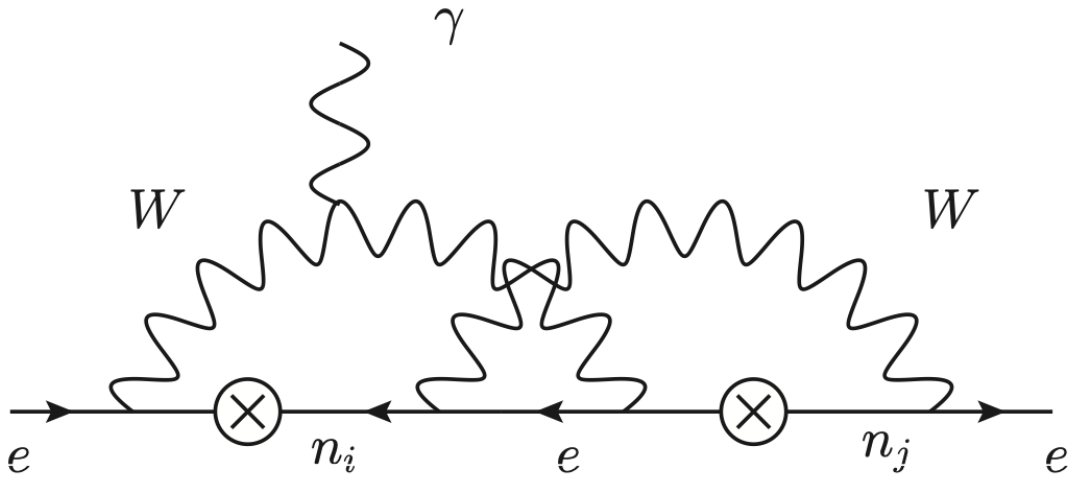}
    \vspace*{-2.5cm}
	\caption{We exhibit two characteristic EDM contributions induced by new physics scenarios. On the left, the diagram (reproduced from \cite{McKeen:2012av}) generates an EDM from $CP$-odd contributions to the Higgs vertex with two photons, $h(F\widetilde F) $. The shaded circle can be resolved in some models as a top quark loop, while the dashed line representing a virtual Higgs will contain scalar-pseudoscalar mixing in multi-Higgs models \cite{Barr:1990vd}.  On the right, the diagram (reproduced from \cite{LeDall:2015ptt}) generates a contribution to the electron EDM from $CP$-odd phases in the neutrino portal and Majorana neutrino masses. This non-planar diagram does not have an analogue in the quark sector.}
	\label{fig:Higgs}
\end{figure}

\subsection{EDMs in the LHC era, and beyond}

Over the past decade and a half, despite a comprehensive exploration of the TeV scale by the LHC experiments, the lack of evidence to date for new BSM phenomenology has necessarily led to a relaxation of theoretical priors on scenarios for high-scale new physics. Prior to the Higgs discovery, supersymmetry (SUSY) was for some time the prevailing theoretical framework for weak-scale new physics, motivated in part by its ability to remove the tuning apparent in the  hierarchy between the weak scale and the Planck scale. Even if SUSY were to form part of the model of high-scale fundamental physics, the discovery of the Higgs boson with a mass of 125\,GeV already greatly reduced the prospects that the LHC experiments would directly produce and observe supersymmetric partners of the SM particles. This conclusion follows from the necessity of a very large renormalization of the tree-level SUSY Higgs mass prediction that obeys $m^{(0)}_h\leq m_Z$. For corrections to be this large, $\delta m^{(1)}_h \geq 34$\,GeV, it is necessary to place the characteristic superpartner mass scale $m_{\rm SUSY}$ above 10\,TeV, as the radiative correction to the Higgs mass, $\delta m^{(1)}_h$, is only logarithmically sensitivity to this scale. SUSY models with $m_{\rm SUSY}\sim100$\,TeV resurrect to some extent the problem of tuning for $v_{\rm EW}$, and weaken the argument that supersymmetry solves the gauge hierarchy problem. Even accepting a certain amount of fine tuning, this Higgs mass consideration turns out to be a powerful argument in favor of very large $m_{\rm SUSY}$ which necessarily has implications for EDMs.

If one persists with supersymmetric models, placing the super-partner masses above 10\,TeV significantly softens the so-called SUSY flavor and $CP$ problems. These are theoretical problems of EW-scale supersymmetry related to the need to suppress flavor-changing and $CP$-violating couplings in the superpartner sector. Heavy superpartners need not conform to flavor alignment \cite{Nir:1993mx} with the SM, and indeed can allow for significant couplings of {\em e.g.} third generation superpartners with the first generation of SM fermions. Similar diagrams, but without an external photon, give SUSY threshold contributions to the masses of the light fermions, $\delta m_f$. Schematically, we can write one-loop SUSY contributions to the electron EDM and electron mass correction in the following form \cite{McKeen:2013dma},
\begin{align}
\delta m_e &\simeq \frac{\alpha}{24\pi\cos^2\theta_W} \times \theta^2_{13} m_\tau \tan\beta, \\
d_e &\simeq \frac{e(\delta m_e)}{5}\times \frac{\sin(\phi_{\rm SUSY})}{m_{\rm SUSY}^2}.
\end{align}
These formulae follow from neutralino (the partner of the SM $\gamma$ and $Z$) and scalar electron one-loop diagrams under the assumption that all SUSY masses are indeed similar, and that there is flavor mixing along the scalar lepton line, $\theta_{13}^2$, that can be sizable but is still treated as a perturbation. The parameter $\tan\beta$ denotes the ratio of the two Higgs vacuum expectation values, $\tan \beta = \langle H_u\rangle/\langle H_d\rangle$, which can lead to an enhancement as large as $O(50)$. There is a significant suppression of $d_e$ due to the fact that in the post-Higgs discovery era the value of $m_{\rm SUSY}$ is very large, but at the same time there can be a parametric enhancement as $\theta^2_{13} m_\tau\gg m_e$ if SUSY flavor alignment is not enforced. If we set $\delta m_e \sim m_e$, we obtain the following sensitivity:
\begin{equation}
\label{desusy}
    d_e\sim 1.4 \times 10^{-28} \,e{\rm cm} \times\left(\frac{\rm 100\,TeV}{m_{\rm SUSY}}\right)^2\times  \left(\frac{\sin(\phi_{\rm SUSY})}{1/\sqrt{2}}\right).
\end{equation}
(If $\delta m_e > m_e$ there will be an additional tuning problem in the electron mass itself, and in that sense $\delta m_e \sim m_e$ is maximal.)
As it stands, this admittedly optimistic estimate suggests that EDMs probe $m_{\rm SUSY}$ above the 100\,TeV scale assuming an anarchical flavor structure among scalar leptons, and a maximal $CP$-violating SUSY phase. The size of $d_e$ in (\ref{desusy}) is already in conflict with the current experimental limits, suggesting that the SUSY flavor/$CP$ problems remain. However, the estimate above neglects possible hierarchies among superpartner masses. The parametrically small gaugino masses $m_{\rm gaugino}/m_{\rm slepton} \sim 0.01$ advocated in Refs.~\cite{Arvanitaki:2012ps,Arkani-Hamed:2012fhg} will suppress $d_e$ by this ratio, bringing it close to current bounds. In addition, if the superpartner sector is taken to be flavor-universal at some high scale, the EDM estimates are somewhat suppressed relative to (\ref{desusy}) \cite{Kaneta:2023wrl}, and typically the sensitivity to $m_{\rm SUSY}$ is limited to tens of TeV. 

In general, a nonzero EDM detection would clearly provide a boost to prospects for finding new physics at the scale of 10's of TeV, and in turn add further motivation for the next generation energy frontier collider. While observable EDMs can of course be generated by new physics at even higher scales, or indeed at sub-EW scales as we discuss in the next subsection, many models inducing $d_e$ close to the current bounds and employing two-loop Barr-Zee type mediation are in reach of current proposals for the next generation of energy frontier colliders \cite{Homiller:2024uxg}.

\subsection{EDMs from dark sectors}

Dark sectors (DS), or neutral BSM particles coupled to the SM via portal interactions (\ref{eq1}), are an example of ``minimal" BSM physics. The focus is on addressing empirical puzzles of the SM, such as the nature of dark matter and neutrino mass, without a direct restriction on the new physics mass scale. Some of these particles, such as axions and heavy neutral leptons are well motivated by providing a solution to the strong $CP$ problem and, even more appealing, by the minimal realization of neutrino mass and mixing. Here we review possible sources of EDMs from DS $CP$-violation. We start from an EFT perspective by listing the lowest dimension portal interactions with the SM that are allowed without modifying the electroweak sector. This involves couplings to new dark $U(1)'$ vectors $A'_\mu$, dark singlet scalars $S$, dark singlet fermions $N$ and, at dimension five, axion-like pseudoscalar particles $a$ (ALPs). Where possible, we add complex phases to these interactions that might generate EDMs.
\begin{eqnarray}
    {\rm Explicitly\,}CP\,{\rm conserving}:&& ~~ A'_{\mu\nu}B_{\mu\nu}; ~~(AS+\lambda S^2)(H^\dagger H);~~aB_{\mu\nu}\widetilde B_{\mu\nu};~~ \partial_\mu a (\bar \psi_i \gamma_\mu\gamma_5 \psi_i).\\
    {\rm Possibly\,}CP\,{\rm violating}:&& ~~ e^{i\phi_N}NLH;~~aB_{\mu\nu} B_{\mu\nu};~~e^{i\delta_{ij}}\partial_\mu a (\bar \psi_i \gamma_\mu\gamma_5 \psi_j),~i\neq j.
    \label{darkCP}
\end{eqnarray}
In these expressions $B_{\mu\nu}$ is the SM hypercharge field strength, $L$ is the left-handed $SU(2)$ doublet, and $\psi_i$ is a generic SM fermion. While at $dim\leq 4$ this list is exhaustive, only a small subset of possible $dim=5$ interactions involving the axion is shown above. Notice that a hypothetical $CP$-odd operator $\widetilde F'_{\mu\nu}B_{\mu\nu}$, in the absence of magnetic charges, can be dropped as a total derivative. In addition to these portal interactions, one can have more interactions among members of the dark sector, possibly with further sources of $CP$ violation ({\em e.g.} $S\bar N i \gamma_5 N$). 

Couplings of the form $e^{i\phi_N}NLH$ are the most well motivated, as their inclusion allows for either Dirac of Majorana masses for active neutrinos contained in the lepton doublet $L$ \cite{Weinberg:1979sa}. The $CP$-violating phases will permeate the neutrino mass matrix and mixing, resulting in $CP$ asymmetries that develop for neutrino flavor transitions propagating over long distances. Additional phases in the neutrino sector can, in fact, generate charged lepton EDMs such as $d_e$ at two-loop EW$^2$ order \cite{Ng:1995cs,Archambault:2004td,Abada:2015trh}. Fig.\,\ref{fig:Higgs}, right side, shows a qualitatively new non-planar diagram that arises in the presence of Majorana masses for $N$. Such diagrams avoid the theorem of vanishing EDMs at two loop order proven for the CKM phase \cite{Shabalin:1978rs} provided that there are at least two such states $N$. Unfortunately, the estimated size of the EDM is quite small. The assumption of a generic heavy $N$ mass pattern with a see-saw mass relation for the active neutrinos limits $d_e$ to very small values \cite{Archambault:2004td}:
\begin{eqnarray}
    d_e \leq e\left(\frac{G_F}{16\pi^2}\right)^2 \times m_em_\nu^2\times 10\log\left(\frac{m_N}{m_W}\right)\; < \;10^{-43} \,{e\rm cm},
    \label{deDS}
\end{eqnarray}
where $m_\nu$ is the mass of the active neutrino states. Note that the suppression by $m_\nu^2/m_W^2$ is very similar to the suppression of {\em e.g.} $\mu\to e\gamma$ and other charged lepton flavor violating amplitudes in this minimal model. 
If the $N$ sector is extended by additional symmetries and cancellations in $m_\nu$, the  mass scale of $m_N$ can be lowered, and the active-sterile mixing angles increased, resulting in a substantial enhancement of $d_e$, up to $\sim10^{-33} \,{e\rm cm}$ \cite{Archambault:2004td,Abada:2015trh,LeDall:2015ptt}, which is larger than the SM CKM expectation for $d_e^{\rm equiv}$ \cite{Ema:2022pmo}, but still considerably smaller than the current experimental sensitivity. One concludes that $d_e$ provides inferior sensitivity to this class of models as compared to flavor-changing processes like $\mu\to e\gamma$ that occur at $O(G_F)$. 

One common feature of the $CP$-violating sources in Eq.~(\ref{darkCP}) is that they necessarily involve particles at or above the weak scale; namely $W,Z,h$ or new BSM states that UV complete ALP interactions. Therefore, the EDM results will always be suppressed by powers of the electroweak scale $\sim v_{\rm EW}$ (or by $f_a \geq v_{\rm EW}$) in the denominator (as in (\ref{deDS})). On the other hand, it is well known that the so-called dark photon portal, $-\frac12\varepsilon F'_{\mu\nu}F_{\mu\nu}$, often generates phenomenology at scales commensurate with the dark photon mass $m_{A'}$, {\em e.g. } below the GeV scale. It is natural to ask whether sub-weak scale DS states can generate measurable EDMs thereby avoiding suppression by factors of $G_F$.

Introducing a dark sector with a dark fermion charged under the new $U(1)'$ group, the dark photon portal, and $CP$-violation residing in a DS interaction of a dark scalar and dark fermion, ${\cal L}_{DS,tree}=\lambda_S S \bar\psi \psi + \lambda_P S \bar\psi i\gamma_5\psi $, one can create a mechanism that transmits $CP$-violation without making any use of weak scale physics. This mechanism involves four loops, and is proportional to the $\lambda_S\lambda_P\alpha^2\alpha_{\rm dark}^2 \varepsilon^4$ combination of couplings. 
For concreteness, we assume $m_S \sim {\rm MeV} \ll m_V \ll m_\psi \sim {\rm GeV}$.
Integrating out the dark fermion $\psi$ at one loop will result in an effective Lagrangian that couples $S$ to $F_{\mu\nu}^{'2}$ and $\widetilde F'_{\mu\nu}F'_{\mu\nu}$. Further integrating out $A'_\mu$ and taking into account mixing with the SM, results in $S\bar e i\gamma_5 e$ and $S\bar p p$ effective interactions, 
\begin{align}
    {\cal L}_{1-loop} = \frac{\alpha_{\rm dark}}{4\pi} \left(\frac23 SV_{\mu\nu}^2+ S\widetilde V_{\mu\nu}V_{\mu\nu}\right)~~\to ~~ {\cal L}_{2-loop} \simeq \frac{3\lambda_P\alpha_{\rm dark}\alpha\varepsilon^2m_e}{2\pi^2m_\psi}\times \log\left(\frac{m_\psi}{m_{A'}}\right)\times  S\bar e i\gamma_5 e+ \frac{2\lambda_S\alpha_{\rm dark}\alpha\varepsilon^2(\pi m_p^{EM}/\alpha)}{3\pi^2m_\psi}\times  S\bar p p,
\end{align}
where $m_p^{EM}\sim 0.6$\, MeV is the proton electromagnetic mass shift. Finally, the exchange by $S$ generates a contribution to $C_S$:
\begin{align}
    {\cal L}_{4-loop} &= C_S^p \frac{G_F}{\sqrt{2}}\bar e i\gamma_5 e\bar p p;~~
    C_S^p\frac{G_F}{\sqrt{2}}  = \frac{1}{m_S^2}\times \frac{m_e(\pi m_p^{EM}/\alpha)}{m_\psi^2}\times \frac{\lambda_S\lambda_P\alpha^2\alpha_{\rm dark}^2 \varepsilon^4}{\pi^4}\times \log\left(\frac{m_\psi}{m_{A'}}\right).
\end{align}
Despite the suppression by the loop factors and coupling constants, the result can be somewhat enhanced for small $m_S$, that can be as low as a few MeV. Accounting for existing limits on portal masses and couplings, and setting $m_S=5$\,MeV, $m_{A'} = 30$\,MeV, $m_\psi=1$\,GeV, $\varepsilon = 10^{-3}$, $\alpha_{\rm dark} =0.5$, and $\lambda_S=\lambda_P=1$, we arrive at a maximal estimate of $C^p_S \sim 4\times 10^{-13}$, which is about three orders of magnitude lower than the current sensitivity. One concludes that sub-GeV dark sectors, utilizing the lowest dimension portal couplings to the SM, do not provide efficient mechanisms for generating large EDMs. This is due either to the necessity for mediation by weak scale charged states, to the high loop order, and/or the high power of small portal couplings (see also \cite{LeDall:2015ptt,Okawa_2019} for further examples). 

One very simple and efficient source of $CP$-violation may be associated with the explicit breaking of the Peccei-Quinn symmetry, that introduces an artificial tilt to the effective axion potential. Recalling that the typical form of the PQ potential with $U(1)_{\rm PQ}$ global symmetry is 
$ \lambda(\Phi^*\Phi - f_a^2)^2$, soft explicit breaking of this potential can be represented as $m_{\rm soft}^2\Phi\Phi + (h.c.)$, where $\Phi$ takes the form $f_a\exp(i a/f_a)$ after PQ symmetry breaking. This correction introduces a linear forcing term, shifting $a$ from its QCD minimum, so that generically $\theta_{\rm ind} \propto m_{\rm soft}^2/m_a^2$. Therefore one should expect that in this scenario the main consequence of explicit PQ symmetry breaking is a nonzero $\bar \theta$, with the corresponding pattern of EDMs discussed in this review. 

Finally, we note that neutral dark sector particles provide an interesting possibility for the enhancement of $d_n$ relative to other EDMs if there exists a ``dark neutron" $n'$ with $CP$-violating mass mixing and a parametrically small mass splitting $\Delta m = m_{n'}-m_n$. Due to the composite nature of neutrons, a mass mixing term $(\delta m)\bar n n'$ will be suppressed by the ratio of the QCD momentum scale and the UV completion scale, {\em e.g.} $\delta m \propto (300\,{\rm MeV})^3/{\rm TeV}^2$. Introducing scalar and pseudoscalar mass mixing terms with real coefficients, and evaluating the $n-n'-n$ diagram at leading order in $(\Delta m)^{-1}$, one finds
\begin{eqnarray}
    {\cal L}_{nn'} = (\delta m)_S\bar n n' + (\delta m)_P\bar n i \gamma_5 n' +(h.c.) ~~\to~~ d_n \sim \mu_n \times \frac{(\delta m)_S(\delta m)_P}{m_n\Delta m } \sim 10^{-26}\,e{\rm cm} \times \left(\frac{1\,\rm MeV}{\Delta m}\right),
    \label{mirrordn}
\end{eqnarray}
where we made use of maximal mixing and maximal $CP$ violation, $(\delta m)_S(\delta m)_P\sim ({300\,\rm MeV})^6/{\rm TeV}^4$.
If $\Delta m\ll {\rm MeV}$, $d_n$ can be enhanced, while all other EDMs will not be impacted at the same level. This is because, {\em e.g.} the neutron energy shifts inside a nucleus are $O(5\,{\rm MeV})$ or larger, and this scale will replace $\Delta m$ in (\ref{mirrordn}) for nuclear EDM and Schiff moment calculations, singling out $d_n$ for $\Delta m \ll 1$\,MeV. Also, one expects the $d_n\neq0, d_p=0$ pattern for nucleons. Admittedly, a parametrically small $\Delta m$ amounts to a certain fine tuning, and this model is further challenged by cosmological and astrophysical constraints \cite{McKeen:2020oyr}.

\section{Conclusions}
\label{Sec5}

Searches for EDMs are an important probe of new fundamental $CP$-violating physics, a critical ingredient in resolving the origin of the matter-antimatter asymmetry. In this review, we have surveyed the theoretical input used to interpret experimental EDM searches and a variety of new physics scenarios that could generate measurable EDMs. Continuing advances in precision have made EDM observables valuable in the exploration of the electroweak frontier, as they have unique sensitivity to flavor-conserving $CP$-odd couplings of $W, Z, h$ and $t$. Models that induce such interactions typically contain additional BSM particles charged under the SM groups which generically allow for $CP$-violating interactions. As one prominent application, the sensitivity of current EDM observables to TeV scale new physics, even when EDMs are generated only at the two-loop level, exerts pressure on viable models of electroweak baryogenesis, excluding a large fraction of model-space. The predominant pre-LHC paradigm for new high-scale physics, weak-scale supersymmetry, is now generically pushed to scales above 10\,TeV. Current EDM sensitivity reaches these scales, and may extend as high as 100\,TeV and above in cases of generic flavor non-universality. At this moment in time, EDMs are one of the few tools available to probe such high-scale models of physics beyond the EW scale. We also reviewed the implications of current EDM sensitivity for new physics in light dark sectors ({\em e.g.} dark photons, sterile neutrinos, dark Higgs bosons) and showed that, except for a few marginal cases, large EDMs are not induced, signaling that a future EDM discovery is more likely to be an indication of new physics at and above the TeV scale. The generic exception to this conclusion would be a small but still dominant $\bar \theta$ source, and therefore it would be important in such a scenario to firmly test the expected ``$\theta$-pattern" across a variety of EDM observables.

The scope of this review did not extend to the experimental EDM landscape, but this is currently very active and a number of current EDM experiments are taking data, while several more in development and close to the data-taking stage (see e.g. \cite{Alarcon:2022ero}). The sensitivity to paramagnetic EDMs \cite{Roussy:2022cmp,ACME:2018yjb} has advanced dramatically over the last twenty years by a factor of $\sim 400$. Assuming $1/\La^2$ scaling in characteristic energy, the corresponding factor of $\sim 20$ gain in sensitivity to $CP$ violation in BSM models means that this jump (albeit indirect) is comparable to the transition from the Tevatron to the LHC. Technological possibilities exist to improve sensitivity in all three categories of EDMs with improvements in existing experiments and methods, and with the development of completely new techniques. Future approaches may be based on novel strategies for assembling, cooling and storing (potentially radioactive) molecules and ions for the purpose of EDM experiments, extending such techniques beyond ThO and HfF$^+$ sources to diamagnetic systems; the potential use of beams of charged particles with ``frozen" spin in storage rings \cite{Farley:2003wt}; the use of more exotic nuclei with the octupole deformations that enhance Schiff moments, as well as employing quantum sensing technology \cite{Auerbach:1996zd,Vutha:2017pej,Arrowsmith-Kron:2023hcr,DeMille:2024djy}. 

On the theoretical front, as indicated above, there remain several areas in the EFT network that require enhanced precision to fully exploit current and future EDM measurements. These include the dependence of nucleon EDMs $d_{n,p}$ and pion-nucleon couplings $\bar{g}^{(0,1)}$ on fundamental quark-gluon $CP$-odd sources ($\bar\th, \tilde{d}_q$), the dependence of Schiff moments $S(d_{n,p},\bar{g}^{(0,1)}$) of large diamagnetic atoms and molecules on the hadronic parameters, and indeed the impact of nuclear polarization effects on $C_S(d_{n,p},\bar{g}^{0,1)})$. The latter case reflects the potential for paramagnetic systems to probe hadronic sources of $CP$-violation. In particular, progress in these areas will allow elucidation of the $\theta$-pattern of EDMs, and more clarity around the reach in energy scale of the neutron EDM and diamagnetic probes. At the hadronic level, {\em ab-initio} lattice QCD calculations, e.g. for $d_{p,n}(\bar\theta,\tilde{d}_q)$, are necessary for systematically improving precision, and several groups continue to tackle this difficult problem, building on the successful analysis of the dependence on quark EDMs. Demonstrating the $d_{n,p}\propto m_*\bar\theta $ dependence remains a challenge. For higher-dimensional operators, complex renormalization and operator mixing challenges also need to be addressed, but progress, e.g. with $\bar g_{\pi NN} (\tilde d_q)$, would greatly enhance the sensitivity of diamagnetic EDM experiments.

\begin{ack}[Acknowledgments]%
 MP is supported in part by the DOE grant \#DESC0011842, and AR is supported in part by NSERC, Canada.
\end{ack}

\seealso{Several comprehensive and topical reviews on EDMs:\\
Comprehensive Reviews \cite{Khriplovich:1997ga,Ginges:2003qt,Pospelov:2005pr,Engel:2013lsa}.}\\
Schiff Moments \cite{Engel:2025uci}\\
Lattice QCD and the neutron EDM \cite{Liu:2024kqy}\\
Global Analysis of Constraints \cite{Degenkolb:2024eve}\\
EDMs and Electroweak Baryogenesis \cite{Morrissey:2012db,vandeVis:2025efm}\\
Experimental status \cite{Alarcon:2022ero}}

\bibliographystyle{Numbered-Style} 
\bibliography{refs}

\end{document}